\documentclass[a4paper,fleqn,usenatbib]{mnras}
\usepackage[T1]{fontenc}
\usepackage{ae,aecompl}
\usepackage{graphicx}	
\usepackage{amsmath}	
\usepackage{amssymb}	

\title[Synthetic photometry for M and K giants]{Synthetic photometry for M and K
giants and stellar evolution: hydrostatic dust-free model atmospheres and chemical
abundances}

\author[B. Aringer et al.]{
B.~Aringer,$^{1,2}$\thanks{E-mail: bernhard.aringer@oapd.inaf.it}
L.~Girardi,$^{2}$
W.~Nowotny,$^{3}$
P.~Marigo,$^{1}$
and A.~Bressan,$^{4,2}$\\
$^{1}$Dipartimento di Fisica e Astronomia Galileo Galilei,
Universit\`a di Padova, Vicolo dell'Osservatorio 3, I-35122 Padova, Italy\\
$^{2}$Osservatorio Astronomico di Padova -- INAF,
Vicolo dell'Osservatorio 5, I-35122 Padova, Italy\\
$^{3}$Department of Astrophysics, University of Vienna,
T\"urkenschanzstra{\ss}e 17, A-1180 Wien, Austria\\
$^{4}$SISSA, via Bonomea 265, I-34136, Trieste, Italy}

\date{Accepted 2016 January 25. Received 2016 January 24; in original form 2015 October 26}
\pubyear{2015}

\begin{document}
\label{firstpage}
\pagerange{\pageref{firstpage}--\pageref{lastpage}}
\maketitle

\begin{abstract}
Based on a grid of hydrostatic spherical COMARCS models for cool stars we have
calculated observable properties of these objects, which will be mainly used in
combination with stellar evolution tracks and population synthesis tools. The high
resolution opacity sampling and low resolution convolved spectra as well as bolometric
corrections for a large number of filter systems are made electronically available.
We exploit those data to study the effect of mass, C/O ratio and nitrogen abundance
on the photometry of K and M giants. Depending on effective temperature, surface gravity and
the chosen wavelength ranges variations of the investigated parameters cause very
weak to moderate and, in the case of C/O values close to one, even strong shifts
of the colours. For the usage with stellar evolution calculations they will be treated as
correction factors applied to the results of an interpolation in the main
quantities. When we compare the synthetic photometry to observed relations and
to data from the Galactic Bulge, we find in general a good agreement. Deviations
appear for the coolest giants showing pulsations, mass loss and dust shells, which
cannot be described by hydrostatic models.
\end{abstract}


\section{Introduction}

As more infrared facilities allowing photometric measurements of faint sources become
available, stellar populations in close galaxies are being increasingly observed in the
corresponding passbands. Our direct neighbours, the Magellanic Clouds, have their evolved
red stars almost completely sampled by surveys such as 2MASS \citep{2003yCat.2246....0C},
IRSF \citep{2007PASJ...59..615K} and VMC \citep{2011A&A...527A.116C} in the near infrared
and by SAGE \citep{2006AJ....132.2268M}, S$^3$MC \citep{2007ApJ...655..212B}, AKARI IRC
\citep{2008PASJ...60S.435I} and WISE \citep{2014MNRAS.442.3361N} in the mid infrared.
There are also data for M~31 and a few dozens of other galaxies in our vicinity. They
have been partially covered by high resolution near infrared imaging from the HST
\citep{2012ApJS..198....6D,2012ApJS..200...18D}. The amount of available photometry for
distant stellar populations will increase a lot, when new space telescopes like JWST or
WFIRST start to operate.

For the more massive galaxies observed in the infrared a large fraction of the detected
bright point sources are cool M stars with very red colours. In general, these objects are
situated in the HRD on the asymptotic giant branch (AGB), in the region of red supergiants
(RSG) and in metal-rich environments often close to the tip of the red giant branch (RGB).
They may be confused with carbon stars, if only broad band photometry is available
\citep{2013ApJ...774...83B}. In order to study their behaviour in colour-colour and
colour-magnitude diagrams of galaxies, one needs spectral libraries to convert stellar
isochrones into the corresponding observable properties. These should cover the relevant
ranges in effective temperature, surface gravity and chemical composition.

Libraries of synthetic spectra have the advantage that they may cover a wide range
of well-defined stellar parameters. There are several grids of hydrostatic models, which
can be or could be used to study K and M giants. Due to the low temperatures of these
objects the opacities caused by many molecular and atomic lines dominate the atmospheres.
Examples are the work of \citet{1989A&AS...77....1B}, \citet{1994A&AS..105..311F} and
\citet{2005ApJ...628..973L} or large libraries based on the MARCS \citep{2008A&A...486..951G}
and the PHOENIX code \citep{2012RSPTA.370.2765A,2013A&A...553A...6H}. For warmer stars one
may also take the spectra from \citet{2003IAUS..210P.A20C}. All of those models cover
different values of the effective temperature, surface gravity, [Fe/H] and in some cases
even [$\alpha$/H]. Results obtained with modified abundances of C, N and O typical for RSG
objects were published by \citet{2007A&A...468..205L}.

In a considerable number of cases hydrostatic models were also used to generate tables
containing synthetic visual and near infrared photometry for cool oxygen-rich giants as a
function of stellar parameters. Examples are the works of \citet{1979A&A....74..313G},
\citet{1989MNRAS.236..653B}, \citet{1992A&A...256..551P}, \citet{1998A&A...333..231B},
\citet{2000AJ....119.1424H}, \citet{2003AJ....126..778V} and \citet{2004AJ....127.1227C}.
The results of \citet{2011ApJS..193....1W} are partly based on observations.

The mentioned hydrostatic models cannot include deviations from spherical symmetry or dynamical
processes like pulsation creating shock waves, and mass loss with dust formation, which become
important in the cooler AGB and RSG objects. A grid for M-type AGB variables covering a limited
parameter range was produced by \citet{2015A&A...575A.105B}. Because of possible uncertainties
of the models, they may be complemented with empirical spectral libraries like the ones from
\citet{1994A&AS..105..311F}, \citet{2000A&AS..146..217L} and \citet[][IRTF]{2009ApJS..185..289R}.

In this work we discuss a grid of hydrostatic COMARCS models for cool stars, which was mainly
designed to provide libraries of synthetic spectra and photometry usable in combination with
the PARSEC \citep{2012MNRAS.427..127B} and COLIBRI \citep{2013MNRAS.434..488M} evolutionary
codes to obtain observable properties of stellar populations. In addition to the primary
parameters effective temperature, surface gravity, mass and metallicity, we have also varied
the abundances of elements like carbon or nitrogen to take the impact of dredge-up processes
into account. This includes new improved calculations for C-type giants, which replace to an
ever-increasing extent the data published by \citet{2009A&A...503..913A}, but are not reviewed
here. In the current text we focus on K and M stars. We first describe the models together with
their synthetic spectra and photometry in Section~\ref{sec_mods}. Then we discuss in
Section~\ref{sec_results} the influence of C/O ratio, nitrogen abundance, mass and the used
water opacity on the atmospheric structures and energy distributions as well as on various
colour indices. In Section~\ref{sec_discussion} the predicted photometric properties are
compared to observed relations and to stars in the Galactic Bulge. Moreover, the application
of the COMARCS data in combination with isochrones is reviewed. The appendix contains a table
with the abundance sets, which are at the moment available in the grid.

\section{Model Atmospheres and Synthetic Spectra}
\label{sec_mods}

\subsection{Opacities and Hydrostatic models}
\label{sec_opac}

In order to study the spectra and photometric properties of K and M
giants we used hydrostatic model atmospheres computed with the COMARCS
program. The latter is based on the version of the MARCS code
\citep{1975A&A....42..407G,2008A&A...486..951G} described in
\citet{1992A&A...261..263J} and \citet{1997A&A...323..202A}.
The models are calculated assuming spherical symmetry and local
thermodynamic and chemical equilibrium (LTE). The continuous
and atomic or molecular absorption at the frequencies used for the
opacity sampling in our calculations is taken from external tables
generated with the COMA program. The formation of dust is not
included. The current approach is identical to the one described
in \citet{2009A&A...503..913A} for a grid of carbon star models.
For further details we refer to this work.

The COMA (Copenhagen Opacities for Model Atmospheres) program is an
opacity generation code that was originally developed by \citet{2000DissAri}
to compute wavelength dependent absorption coefficients for cool model
atmospheres. It has also been used to determine weighted mean opacities for
stellar evolution calculations \citep{2009A&A...494..403L, 2007ApJ...667..489C}.
In combination with an attached radiative transfer module COMA was applied
in a considerable number of investigations to obtain synthetic photometry
and spectra at very different resolution \citep[e.g.][]{2009A&A...503..913A,
2014A&A...566A..95E,2010A&A...514A..35N,2014A&A...567A.143L,2015MNRAS.451.1750U}.
In the central routines the abundances of many important ionic and molecular
species are evaluated assuming chemical equilibrium. Subsequently, the resulting
partial pressures can be used to compute the continuous and line opacity at each
selected frequency point (in LTE). At the moment, up to more than 175 million
transitions are included in a calculation covering the complete COMARCS wavelength
range. This number, which is dominated by molecular species with more than two atoms
(e.g.\ H$_2$O, HCN, CO$_2$) or electronic transitions (e.g.\ TiO, CN) depends on
the chosen line lists. In addition, COMA is also able to produce opacities
for a considerable amount of dust types, if information concerning the degree
of condensation is provided \citep[e.g.][]{2015A&A...575A.105B,2011A&A...529A.129N,
2015ASPC..497..397N}. It should be noted that equilibrium abundances of solid particles
are not computed, since such a situation never appears in astrophysical environments.

More information concerning the COMA code can be found in \citet{2009A&A...503..913A}.
In comparison to that work we have added some new opacity sources mainly important
for the structures and spectra of dwarf stars. Line lists for the hydrides MgH
\citep{2003ApJ...582.1059W,2003ApJS..148..599S}, TiH \citep{2005ApJ...624..988B}
and CaH \citep{2003JChPh.118.9997W} are now taken into account. Their features
appear primarily in the visual range. For CaH and MgH the data for the chemical
equilibrium constants come from \citet{1984ApJS...56..193S} with a revised
dissociation energy of 1.285~eV for MgH \citep{shayesteh_etal07}. The corresponding
values for TiH were taken from \citet{2005ApJ...624..988B}. At this point it should
also be noted that we have changed the dissociation energy of CrH from 2.86 to
2.17~eV \citep{2001JChPh.115.1312B} resulting in significantly lower abundances
(up to two orders of magnitude) and weaker bands of the molecule compared to
\citet{2009A&A...503..913A}.

Other species which have been added to the line opacity calculations are CS
\citep{1995A&AS..114..175C} and CH$_4$. In the case of methane COMA can currently
only process the HITRAN 2008 data \citep{2009JQSRT.110..533R} with a quite limited
number of transitions. This is no problem for the results presented here, since the
hydrostatic model atmospheres do not extend into the cool temperature ranges where
CH$_4$ becomes important. At the moment the following molecules are included
in COMA and the line opacities used for this work: CO, CH, C$_2$, CS, CN,
H$_2$O, C$_2$H$_2$, HCN, C$_3$, SiO, OH, TiO, VO, YO, ZrO, CO$_2$, SO$_2$,
CH$_4$, HF, HCl, FeH, CrH, MgH, TiH and CaH\@.

Line lists from the HITRAN database may be selected in COMA as an option for
CO, H$_2$O, CO$_2$, OH, HF and C$_2$H$_2$, while they represent the only available
source for SO$_2$, HCl and CH$_4$. In the case of OH they are the preferred
choice. In comparison to the computations described in \citet{2009A&A...503..913A},
where only the 2004 version of HITRAN \citep{2005JQSRT..96..139R} could be used,
it is now also possible to take the data from HITRAN 2008 \citep{2009JQSRT.110..533R}.
For the work presented here we extracted the transitions of SO$_2$, HCl and CH$_4$
from HITRAN 2008, while we kept HITRAN 2004 for OH (better agreement with observed
mid IR spectra). In addition, COMA can process the HITEMP 2010 data for H$_2$O,
CO$_2$, OH and CO \citep{2010JQSRT.111.2139R}, which may improve the results for
the first three of the molecules. However, since that option was not available
when the work on the COMARCS grid was started, the HITEMP 2010 lists are not
considered here. This is not a big problem, because for H$_2$O and OH the changes
of the overall absorption and medium or low resolution spectra are small, if one
compares to calculations with the current standard data (HITRAN for OH and BT2
from \citet{2006MNRAS.368.1087B} for water). The update of HITEMP to the 2010
version produces some differences in the CO$_2$ bands around 4.4~$\mu$m. But
even in the coolest COMARCS models the features of carbon dioxide remain always
very weak.

\begin{figure*}
\centering
\includegraphics[width=15.0cm,clip,angle=270]{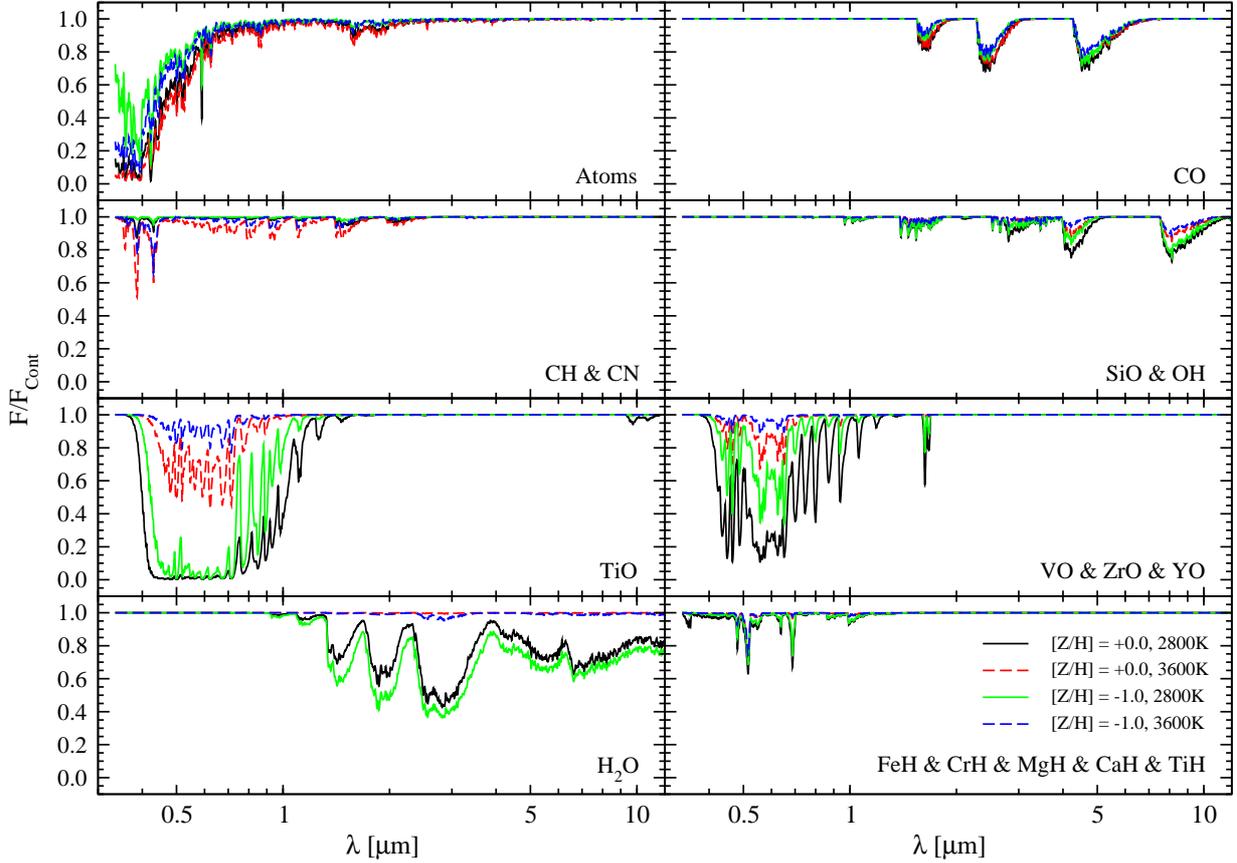}
\caption{Continuum normalized spectra for important molecular species and
atomic lines based on COMARCS models with $\rm log(g~[cm/s^2]) = 0.0$,
$\rm M/M_{\odot} = 1.0$ and scaled solar abundances. Two different
metallicities ($\rm Z/Z_{\odot} = 1.0 \rightarrow [Z/H] = 0.0$,
$\rm Z/Z_{\odot} = 0.1 \rightarrow [Z/H] = -1.0$) and temperatures
($\rm T_{eff} = 2800~K$, full lines, $\rm T_{eff} = 3600~K$, dashed lines)
are shown.}
\label{ari_molec01}
\end{figure*}

It should be mentioned that the changes discussed in the previous paragraphs
do not have a significant effect on the overall opacity and atmospheric structure
in the models. As in \citet{2009A&A...503..913A} Doppler profiles including the
thermal and the microturbulent contribution were assumed for the molecules, while
the atomic transitions are broadened with Voigt functions, if damping constants
are available. In Fig.~\ref{ari_molec01} we show absorption spectra of different
species important for K and M stars, which are calculated from COMARCS atmospheres
with $\rm log(g~[cm/s^2]) = 0.0$, $\rm M/M_{\odot} = 1.0$ and scaled solar
abundances. Four models having solar ($\rm Z/Z_{\odot} = 1.0$) or ten times lower
($\rm Z/Z_{\odot} = 0.1$) metallicity and $\rm T_{eff} = 2800~K$ or 3600~K are
included. The spectra were normalized using a computation without any line
opacities and have a resolution of R~=~200. Species showing similar behaviour
are combined in groups: atoms, CO, CH \& CN, SiO \& OH, TiO, VO \& ZrO \& YO,
H$_2$O, FeH \& CrH \& MgH \& CaH \& TiH\@.

As one can see in Fig.~\ref{ari_molec01}, TiO and atoms dominate the opacities
below 1~$\mu$m. The corresponding absorption increases with higher metallicity
and lower temperature. Especially in cooler objects the stellar radiation is
almost completely blocked in broad regions of the spectrum, which has a severe
impact on the atmospheric structure. The strong heating effect caused by TiO
is discussed in \citet{1997A&A...323..202A}. At low temperatures water becomes
an important absorber in the whole infrared range. For solar abundances this
happens in giants cooler than 3200~K, while in dwarfs intense H$_2$O bands
appear already at around 4200~K, since higher pressures favour the formation
of polyatomic molecules. An interesting trend visible in Fig.~\ref{ari_molec01}
is that the features of H$_2$O get stronger with decreasing metallicity. This
behaviour can mainly be explained by the reduced heating of the atmospheres due
to TiO\@. In order to observe such an effect, stars with identical parameters
($\rm T_{eff}$, log(g)) need to be compared. However, this is not so easy, if one
studies different stellar populations.

In Fig.~\ref{ari_molec01} it is also
obvious that at cooler temperatures there exists no point in the spectra,
which is not affected by intense molecular or atomic absorption. Due to the
impact of line opacities on the atmospheric structures and measurable
pseudo-continua the relation between chemical abundances and the apparent
depth of spectral features may become quite complicated.

\begin{figure}
\includegraphics[width=7.0cm,clip,angle=270]{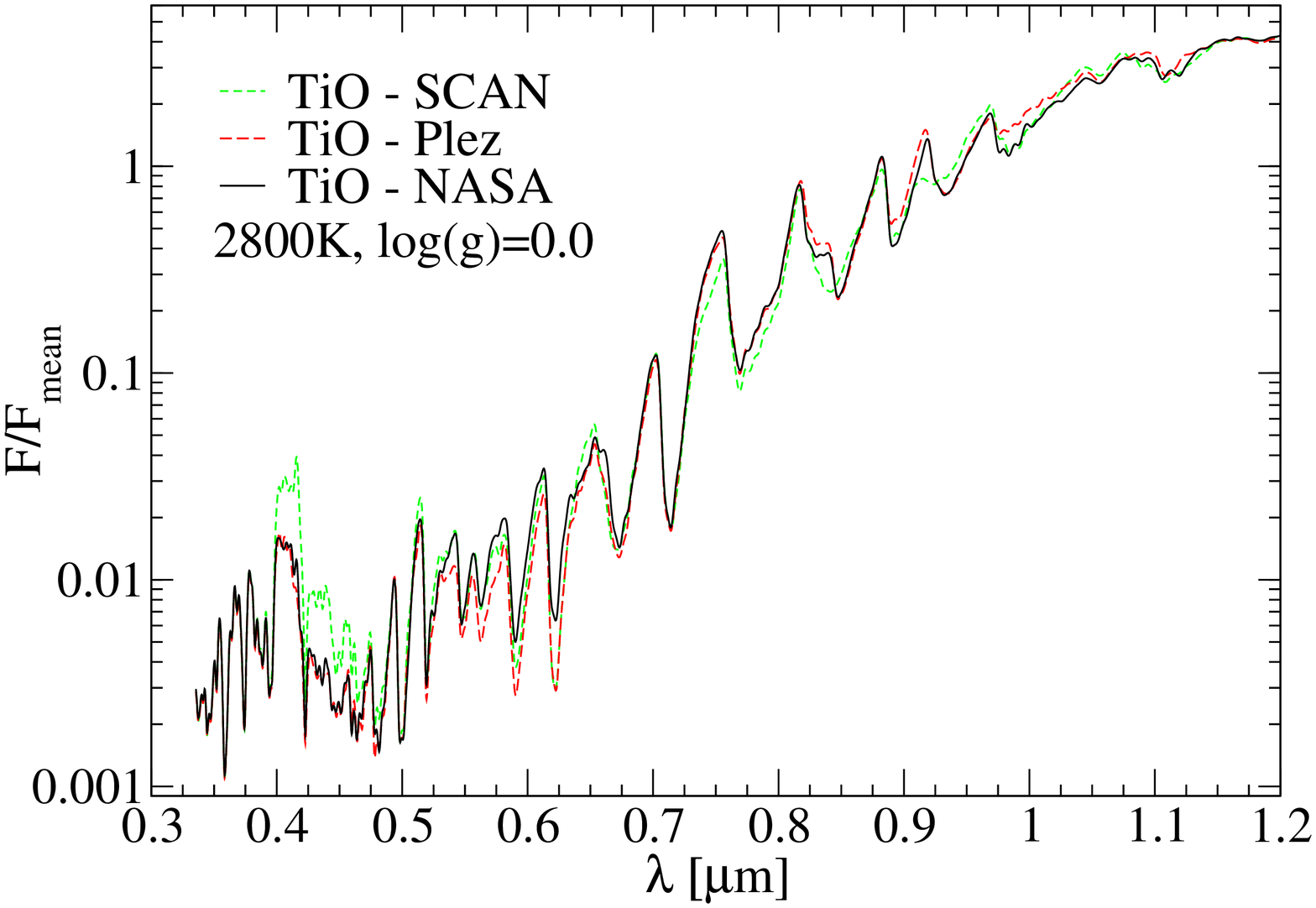}
\caption{Spectra calculated with different TiO data based on a COMARCS
model with $\rm T_{eff} = 2800~K$, $\rm log(g~[cm/s^2]) = 0.0$, one solar mass
and solar metallicity. The NASA-AMES (black, used in this work as standard),
Plez (red) and SCAN (green) line lists are compared. The intensities
$\nu$L$_\nu$ are normalized to their mean value.}
\label{ari_lists1}
\end{figure}

\begin{figure}
\includegraphics[width=7.0cm,clip,angle=270]{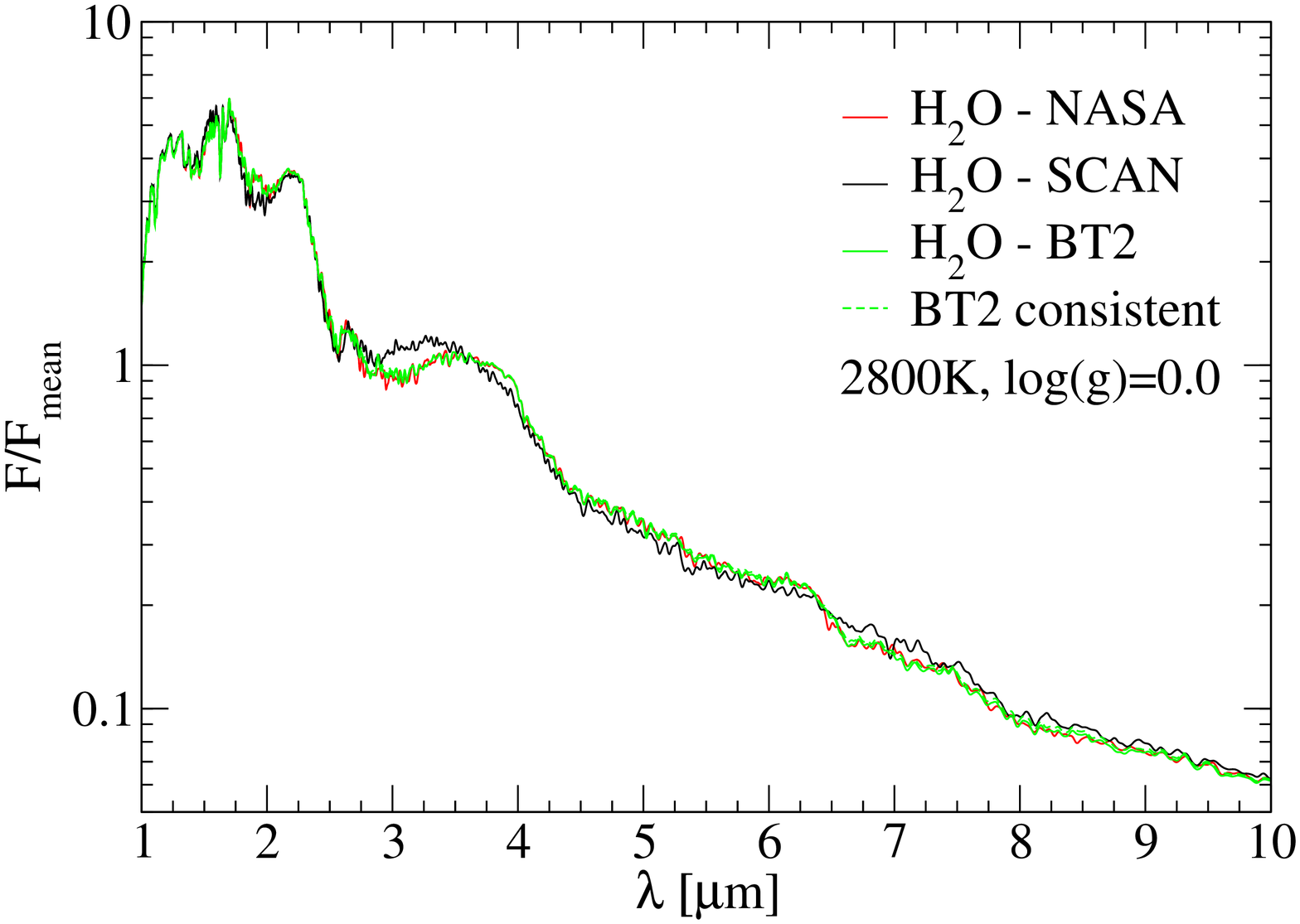}
\caption{Spectra calculated with different H$_2$O data based on a COMARCS
model with $\rm T_{eff} = 2800~K$, $\rm log(g~[cm/s^2]) = 0.0$, one solar mass
and solar metallicity. The SCAN (black, used in this work as standard),
BT2 (green, used in this work as standard) and NASA-AMES (red) linelists
are compared. The intensities $\nu$L$_\nu$ are normalized to their mean value.
We also show the result obtained with a consistent usage of the BT2 data
for the computation of the atmospheric structure and the spectrum (green
dashed), which is very similar to the one, where BT2 was only taken for the
a posteriori radiative transfer (green full).}
\label{ari_lists2}
\end{figure}

There are still some uncertainties concerning the opacity data, especially
for species with a large number of transitions and at higher spectral
resolutions. This can already be seen in Fig.~\ref{ari_kuru} where we compare
the results from different hydrostatic model approaches. The effect of changing
the selected line list for TiO and H$_2$O in COMA is shown in Figs.~\ref{ari_lists1}
and \ref{ari_lists2}. Nevertheless, in the case of the K and M giants studied here
we do not expect that any uncertainties of the used data or possible missing opacities
will have a significant influence on the overall absorption and atmospheric structures.

In Fig.~\ref{ari_lists1} we compare synthetic spectra for a COMARCS model with
$\rm T_{eff} = 2800~K$, $\rm log(g~[cm/s^2]) = 0.0$, one solar mass and solar
metallicity, which were calculated using different TiO data. The computations
presented here are normally done with the NASA-AMES list \citep{1998FaDi..109..321S}.
In addition, we have included results obtained with the Plez \citep{1998A&A...337..495P}
and the SCAN list \citep{1994A&A...284..179J,1997IAUS..178..441J}. The spectra, which
do also contain the features of the other species, cover the wavelength range up to
1.2~$\mu$m at a resolution of R~=~200. It should be noted that the same COMARCS model
calculated with the NASA-AMES data was used in all three cases. This inconsistency
is no problem, since the studied lists produce quite similar overall opacities
and atmospheric structures. There are only some differences on smaller scales.

A similar comparison is done in Fig.~\ref{ari_lists2} where we investigate the
effect of using different H$_2$O data. The stellar parameters of the corresponding
COMARCS model are the same as in Fig.~\ref{ari_lists1}. We present results obtained
with the BT2 \citep{2006MNRAS.368.1087B}, SCAN \citep{2001A&A...372..249J} and
NASA-AMES \citep{1997JChPh.106.4618P} list. HITEMP 2010 \citep{2010JQSRT.111.2139R},
which may be regarded as the most recent and complete collection of water lines, is
not included in the plot. However, at the selected low resolution of R~=~200 the
differences between BT2 and HITEMP 2010 are rather small. It should be noted that
HITEMP is partly based on the work of \citet{2006MNRAS.368.1087B}. The atmospheres
in the COMARCS grid are computed using the SCAN data, which represent therefore
the prime choice for the calculation of the synthetic spectra. For all oxygen-rich
models with temperatures cool enough that water absorption may become important
the a posteriori radiative transfer has also been done involving the BT2 list.
The limit was in most of the cases set to $\rm T_{eff} = 4000~K$, but may become
higher than 5000~K for dwarfs. In the plots of this paper we show usually the
results obtained with the BT2 data if available. Like for TiO, the inconsistent
treatment of opacities in the calculation of COMARCS atmospheres and synthetic spectra
will not cause problems, since the different lists give quite similar structures.
It should be noted that this is true for hydrostatic models and the range of
parameters covered by the COMARCS grid. If a significant part of the computed atmosphere
extends to gas temperatures below 1000~K, SCAN and BT2 produce deviating results.
The reason for this is a systematic difference in the overall absorption above
6~$\mu$m as one can see in Fig.~\ref{ari_lists2}. A discussion of the behaviour
of SCAN and NASA-AMES in the mid infrared and a comparison to ISO-SWS observations
are presented in \citet{2002A&A...395..915A}.

In Fig.~\ref{ari_lists2} we compare two different spectra obtained with the
BT2 line list. The first one is based on a standard COMARCS atmosphere, which has
been calculated with the SCAN data. The second energy distribution results from a
consistent usage of the BT2 opacities for the determination of the model structure
and the a posteriori radiative transfer. It is obvious that the differences between
the two spectra remain in all regions very small. This conclusion is also valid, if
one plots the absolute intensities without normalization. The relative deviations of
the frequency integrated radiative fluxes in the various layers up to the surface
caused by the inconsistent treatment of the water opacity in the first case do never
exceed values around 2~\%. This confirms that our approach concerning the BT2 spectra
is not too problematic.

In addition to the new and updated line lists we have also revised the continuous
opacities by including the collision induced absorption (CIA) of H$_2$~--~H$_2$
\citep{2001JQSRT..68..235B,2002A&A...390..779B}, H$_2$~--~H \citep{2003A&A...400.1161G},
H$_2$~--~He \citep{2000A&A...361..283J} and H~--~He \citep{2001ApJ...546.1168G}. This
is especially important for the very cool and dense dwarf atmospheres at low metallicity,
while it has no effect on the structures and spectra of the giants studied here.

\subsection{Model parameters}
\label{sec_par}

The COMARCS grid, which was mainly designed to transfer the results of stellar
evolution calculations into observable properties of the various objects
\citep[e.g.][]{2009A&A...503..913A}, aims at cool stars. The typical range
of effective temperatures covered by the models is between 2500 and 5000~K
with values of $\rm log(g~[cm/s^2])$ extending from 5.0 to $-1.0$. Some hotter
($\rm T_{eff}$ up to 7170~K) and more compact ($\rm log(g~[cm/s^2])$ up to 5.32)
computations are also included. An overview of the COMARCS grid can be found in
the appendix in Table~\ref{ari_comarcs}, where we list the opacity sets available
at the moment. They correspond to an input file produced by COMA and are
characterized by a certain selection of elemental abundances, microturbulent
velocity ($\xi$) and source data for the absorption. Each of them may contain
models with different effective temperature, surface gravity or mass. The range
in $\rm T_{eff}$ and log~(g) covered by the various sets is also listed in the
table. Since the COMARCS grid is constantly growing, the information given in the
appendix represents only a snapshot at the time when this paper was produced.
An updated and complete catalogue of the available models may be obtained from
\url{http://starkey.astro.unipd.it/atm}.

The central part of the COMARCS database are several subgrids of different
metallicity where the relative abundances of all elements heavier than He (Z) have been
scaled in the same way. The values with reference to the Sun log~(Z/Z$_{\odot}$) or
[Z/H] range from $-2$ to +1 separated by steps of 0.5. The corresponding solar
chemical composition is based in essence on \citet{2009MmSAI..80..643C,2009A&A...498..877C}
resulting in a C/O ratio close to 0.55. In this work Z/Z$_{\odot}$ and [Z/H] represent
the relative abundance of the bulk of metals. A variation of the amount of one or
more elements like carbon or nitrogen will not change these numbers. Since the
current COMARCS database does not include opacity sets with a special scaling
for iron, [Z/H] is always equal to [Fe/H]. The subgrids of different metallicity
cover effective temperatures between 2600 and 5000~K with steps of 100~K or smaller,
while the values of $\rm log(g~[cm/s^2])$ range from 5.0 to $-1.0$ with a maximum
increment of 0.5. Models with $\rm log(g~[cm/s^2]) < 0.0$ were usually only computed
for stars cooler than about 3500~K\@. The standard microturbulent velocity of the
calculations was set to $\xi = 2.5$~km/s, which is in agreement with our previous
work \citep{2009A&A...503..913A,1997A&A...323..202A} and high resolution observations
of M giants \citep[e.g.][]{1985ApJ...294..326S,1990ApJS...72..387S}. For other
groups of objects covered by the COMARCS grid different typical $\xi$ values will be
appropriate. An example are dwarfs or red supergiants, which are not discussed in
this paper. However, the influence of the microturbulence on the atmospheric structure
and low resolution spectra remains for most of the cooler models quite limited, if the
deviations do not become too large. This is due to the huge number of weak overlapping
lines that dominate the opacity. For solar metallicity the COMARCS database includes
also calculations with $\xi = 1.5$ and 5.0~km/s. In general, the model atmospheres in the central
subgrids were computed assuming the mass of the Sun. Especially for the more extended
objects with $\rm log(g~[cm/s^2]) \le 0.0$, where sphericity effects may become
important, different selections of this parameter have been considered.

Concerning the evaluation of the convective fluxes we follow the approach described
and applied by \citet{1975A&A....42..407G,2008A&A...486..951G} for their MARCS models as
we use the mixing length formalism from \citet{1965ApJ...142..841H} with the same
parameters ($\alpha = 1.5$, $\nu = 8.0$, y~=~0.076). For the calculation of the turbulent
pressure we have adopted $\beta = 1.0$. The velocity of the convective elements from the
equations was taken. According to \citet{2008A&A...486..951G} this may result in values
being too high. However, the influence of the chosen turbulent pressure on our synthetic
spectra and photometry is limited, since only the innermost layers are affected.

\begin{figure}
\includegraphics[width=7.0cm,clip,angle=270]{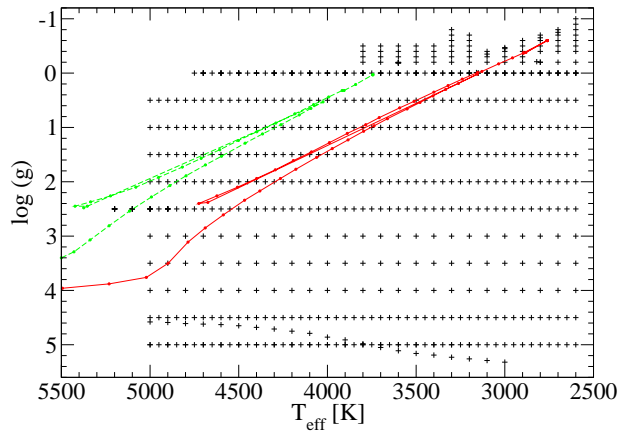}
\caption{Values of effective temperature and $\rm log(g~[cm/s^2])$ covered
by the current COMARCS grid in the case of solar abundances (crosses).
Models of different mass are not distinguished. In addition, evolutionary
tracks for stars with one solar mass and $\rm Z/Z_{\odot} = 1.0$ (full line,
red) as well as $\rm Z/Z_{\odot} = 0.1$ (dashed line, green) are displayed.}
\label{ari_grid}
\end{figure}

In Fig.~\ref{ari_grid} we show the subgrid of $\rm T_{eff}$ and log~(g) values
for the solar chemical composition. Models of different masses and microturbulent
velocities are included without any further distinction. Up to 3800~K the
calculations have been extended to $\rm log(g~[cm/s^2]) < 0.0$. The high density
of grid points in that region and constraints of the lower limit for the surface
gravity around 3000~K are partially caused by convergence problems
\citep{2009A&A...503..913A}. In the domain of the compact objects there is a
sequence of COMARCS atmospheres with effective temperatures from 5000 to 3000~K, which
crosses the $\rm log(g~[cm/s^2]) = 5.0$ line. This corresponds to the ZAMS
of dwarfs having solar chemical composition as derived from the PARSEC code
\citep{2012MNRAS.427..127B} for stellar evolution.

Figure~\ref{ari_grid} contains also two sequences of models connected by lines,
which have been placed along evolutionary tracks for stars with the mass of the
Sun. The corresponding values of $\rm T_{eff}$ and log~(g) were computed using
PARSEC and for the latest stages the COLIBRI code \citep{2013MNRAS.434..488M}.
They cover the whole period between the ZAMS and the final phase of the AGB\@.
The sequences, which extend to much higher temperatures than the rest of the
shown COMARCS subgrid (5940 \& 7170~K), differ by their chemical composition.
The first one is compatible with the other models in the figure, since it has
also solar abundances. The second track, which is considerably hotter, was
calculated assuming [Z/H]~=~$-1$. Because of intense stellar winds at the
late stages of the AGB evolution that are included in the COLIBRI computations,
the mass of the last few objects in the sequences is reduced (0.74 \& 0.79~M$_{\odot}$
at the final point). We have taken this effect into account when producing the
COMARCS models.

As one can see in Table~\ref{ari_comarcs}, the COMARCS database contains also opacity
sets where the abundances of individual elements have been changed separately. The
most important example is the amount of carbon, which is specified here by the C/O
ratio. This varies during the late stages of stellar evolution due to several
nucleosynthesis and dredge-up processes \citep[e.g.][]{1999ARA&A..37..239B,
2011ApJS..197...17C}. The data include a number of subgrids and sequences computed
with C/O values different from the one associated with the solar composition (0.55),
which cover the interval between 0.25 and 10.0. Models were usually produced for
those combinations of effective temperature and surface gravity where the
corresponding intrinsic modifications of the abundances are expected. As an
example, there was no need to take dwarfs into account. This results in subgrids
much smaller than the ones with a metallicity scaled solar composition. The opacity
sets for carbon stars (C/O~$>$~1) are listed in the end of Table~\ref{ari_comarcs}.
The models therein will replace the database of \citet{2009A&A...503..913A}.

The COMARCS data allow also to investigate the effect of an increased nitrogen
abundance. Models with [N/Z]~=~+1 exist for two different metallicities:
[Z/H]~=~0 and $-1$. At the moment sequences with $\rm log(g~[cm/s^2]) = 0.0$ and
+2.0 covering temperatures between 2600 and 5000~K as well as COMARCS atmospheres placed
along an evolutionary track of a star with one solar mass (see Fig.~\ref{ari_grid}
\& text above) are available. The effects of enhancing nitrogen and changing C/O
will be discussed in the next sections.

\subsection{Synthetic Spectra and Photometry}
\label{sec_syn}

For the complete grid of hydrostatic COMARCS atmospheres we computed synthetic
opacity sampling spectra covering the range between 400 and 29814.7~cm$^{-1}$
(0.3354 to 25.0~$\mu$m) with a resolution of R~=~10000. Following the approach
described in \citet{2009A&A...503..913A} or \citet{2011A&A...529A.129N} for
carbon stars, we used the COMA code in combination with an attached spherical
radiative transfer program \citep{1997A&A...324..617W}. Except for the denser
distribution of frequency points, the treatment of the continuous and line
absorption was in general consistent with the input data for the production
of the models. As it is discussed in Section~\ref{sec_opac}, that is not valid
for the spectra, which have been calculated involving the BT2 list for water.
However, this does not create a severe problem, since the different applied
H$_2$O data result in similar hydrostatic structures.

Because of the statistical nature of the opacity sampling approach, only the
average over a larger number of frequency points (about 20 to 100) ensures a
realistic representation of the observed stellar spectra \citep{2011A&A...529A.129N}.
Thus, we decreased the resolution of the output of the radiative transfer code to
R~=~200 by convolving it with a Gaussian function. We did not include any additional
line broadening induced for example by a macroturbulent velocity. The wavelength grid
has been done with a spacing corresponding to R~=~2000 resulting in a smooth appearance
of the final spectra. The same procedure was already used by \citet{2009A&A...503..913A}
to obtain their low resolution data. Examples for the output can be found in
Figs.~\ref{ari_molec01}, \ref{ari_lists1}, \ref{ari_lists2} \& \ref{ari_kuru}.
The spectra are electronically available as described in Section~\ref{sec_spec}.

We also produced synthetic bolometric corrections for a large set of filters following
the same formalism as described in \citet[][eq.~7 \& 8]{2002A&A...391..195G}. First we
convolved the transmission curve with the calculated opacity sampling spectra. This was
done with an integration of the energy or the number of detected photons depending, if
amplifiers or counting devices were used for the definition of the investigated
photometric system. Subsequently, the obtained magnitudes were scaled with respect
to a reference intensity in order to consider the zero points. This quantity was
determined from a convolution with the filter curve applied to a constant flux per
unit wavelength (frequency) in the case of STmag (ABmag) systems or to a Vega
($\alpha$~Lyr) spectrum for Vegamag data. We took the energy distribution of the
standard star from the work of \citet{2007ASPC..364..315B}. The corresponding values
are expected to be accurate to within about 2\% in the optical and near infrared
range. The latest table of synthetic bolometric corrections is electronically
available as described in Section~\ref{sec_phot}.

\section{Results}
\label{sec_results}

In this section we want to present some results obtained using the COMARCS grid
with special emphasis on the effects of C/O ratio, nitrogen abundance and stellar
mass. The impact of these quantities on the atmospheric structures, spectra and
colours will be discussed. It can be regarded as a correction that has to be
applied to the properties of a star predicted from the main parameters, which
are temperature, surface gravity and overall metallicity (see Section~\ref{sec_evo}).
We investigate, where the mentioned effects become important. As it was already
stated, the values of C/O and [N/Z] change during the late stages of stellar
evolution. If the C/O ratio exceeds one, a severe modification of the atmospheric
structure and spectral appearance happens, since the object turns into a carbon
star. However, this case will not be considered here, because we focus on K and
M giants. The evolutionary tracks for one solar mass shown in Fig.~\ref{ari_grid}
that are used for some plots of this paper do also not reach C/O~$\ge$~1.0. An
enhancement of s-process elements like Y or Zr, which is often connected to an
increased C/O value, is as well not discussed, since we do not cover S stars in
our study. The properties of the latter group of objects are investigated by
\citet{2011ASPC..445...71V}.

\subsection{Atmospheric Structure}
\label{sec_atstru}

\begin{figure}
\includegraphics[width=7.0cm,clip,angle=270]{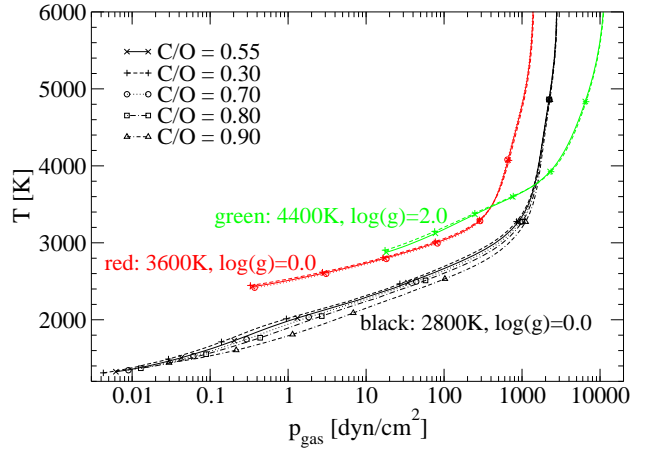}
\caption{The effect of the C/O ratio on the atmospheric temperature
versus gas pressure structure is shown for different COMARCS models
with solar mass and metallicity. Three examples are included:
$\rm T_{eff} = 2800~K$, $\rm log(g~[cm/s^2]) = 0.0$ (black) with
C/O~=~0.30, 0.55, 0.70, 0.80, 0.90, $\rm T_{eff} = 3600~K$,
$\rm log(g~[cm/s^2]) = 0.0$ (red) with C/O~=~0.30, 0.55, 0.70 and
$\rm T_{eff} = 4400~K$, $\rm log(g~[cm/s^2]) = 2.0$ (green) with
C/O~=~0.30, 0.55. C/O~=~0.55 corresponds to the solar value.
The plot symbols placed along the different curves mark steps of
$\rm \Delta log(\tau_{Ross}) = 1$ starting with $-5$ at the outer edge.}
\label{ari_costruc}
\end{figure}

\begin{figure}
\includegraphics[width=7.0cm,clip,angle=270]{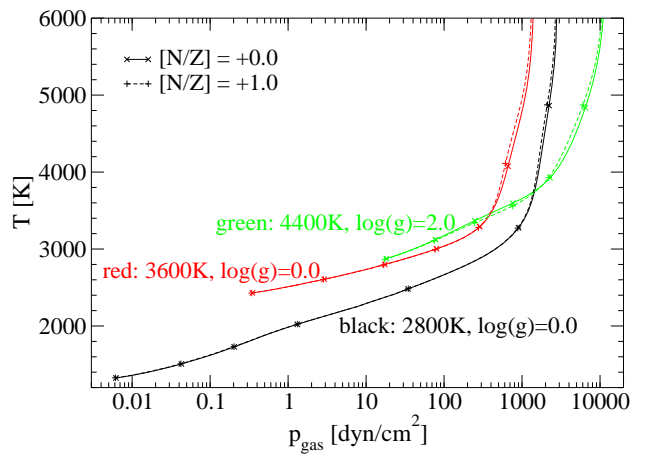}
\caption{The effect of the nitrogen abundance on the atmospheric temperature
versus gas pressure structure is shown for different COMARCS models
with solar mass and metallicity. Three examples are included:
$\rm T_{eff} = 2800~K$, $\rm log(g~[cm/s^2]) = 0.0$ (black),
$\rm T_{eff} = 3600~K$, $\rm log(g~[cm/s^2]) = 0.0$ (red) and
$\rm T_{eff} = 4400~K$, $\rm log(g~[cm/s^2]) = 2.0$ (green).
[N/Z]~=~0.0 (full) corresponds to the solar value of
$\rm log(\varepsilon_N/\varepsilon_H) + 12 = 7.90$ and [N/Z]~=~+1.0
(dashed) to N enhanced by a factor of 10.
The plot symbols placed along the different curves mark steps of
$\rm \Delta log(\tau_{Ross}) = 1$ starting with $-5$ at the outer edge.}
\label{ari_nstruc}
\end{figure}

In Fig.~\ref{ari_costruc} we present the atmospheric temperature pressure structure of
various hydrostatic models taken from the COMARCS grid, which have the mass and overall
metallicity of the Sun. For three different combinations of $\rm T_{eff}$ and log~(g) the
effect of changing the C/O ratio is shown. Values between 0.3 and 0.9 are covered including
0.55 for the solar composition. Models with a high C/O exceeding 0.7 are only available
for the coolest set with $\rm T_{eff} = 2800~K$ and $\rm log(g~[cm/s^2]) = 0.0$. Such a
large relative enhancement of carbon can appear during and after the AGB phase as a
consequence of the third dredge-up \citep{1999ARA&A..37..239B,2011ApJS..197...17C}.
The additional combinations are $\rm T_{eff} = 3600~K$ with $\rm log(g~[cm/s^2]) = 0.0$
and $\rm T_{eff} = 4400~K$ with $\rm log(g~[cm/s^2]) = 2.0$.
The figure shows that there is a clear trend of colder structures with increasing C/O
ratio, but significant changes happen solely above 0.7. The latter may be partly explained
by the fact that the amount of oxygen atoms not bound in CO gets more crucial for the
formation of other species like TiO or H$_2$O, if it becomes very low. However, one should
always keep in mind that the relations can be quite complicated, if molecules important
for the overall opacity are involved. For example, the abundance of TiO, which causes a
strong heating of the atmospheres, decreases in the inner regions of the models having
$\rm T_{eff} = 2800~K$ and $\rm log(g~[cm/s^2]) = 0.0$ at higher C/O values, while it grows
in the outer areas below $\rm T_{gas} = 2400~K$.

In Fig.~\ref{ari_nstruc} we demonstrate, how increasing the nitrogen abundance by a
factor of ten affects the atmospheric temperature pressure structure. This enhancement
can be regarded as typical for the late stages of stellar evolution, since the maximum
values predicted by codes like COLIBRI \citep{2013MNRAS.434..488M} range between about
2 and 100 depending on mass and composition. We show three sets of models with the
same parameters as the ones used in Fig.~\ref{ari_costruc} for the C/O ratio. It is
obvious that the impact of changing [N/Z] remains in general small. It disappears almost
completely for the coolest objects, where the CN bands become weaker and their importance
is additionally reduced by the intense absorption of other species like TiO or water.

\subsection{Spectra}
\label{sec_spec}

The complete grid of our R~=~10000 opacity sampling and convolved low resolution
R~=~200 spectra computed from the COMARCS models can be found in
\url{http://starkey.astro.unipd.it/atm}.
The corresponding tables consist of three columns: wavelength [{\AA}], continuum normalized
flux and specific luminosity times frequency ($\rm \nu \cdot L_{\nu}$~[erg/s]). The second
quantity is obtained from dividing the result of a full radiative transfer calculation by
one where the atomic and molecular line absorption is totally neglected. This allows to
evaluate the apparent intensity of the various spectral features. However, such continuum
normalized data should be interpreted with care, since the optical depth scale changes
when opacities are removed. An example can be seen in Fig.~\ref{ari_molec01}, where we
investigate the contribution of different species to the total absorption. In this case
a calculation including only the corresponding transitions was divided by one without any
lines. The resolution is again R~=~200. We want to emphasize that the R~=~10000 opacity
sampling spectra must not be directly compared to observations as it was explained in
Section~\ref{sec_syn}. The parameters used to characterize the different entries in our
database can be found in Section~\ref{sec_phot}, where we describe the tables with the
synthetic photometry.

In the following sections we study the changes of the R~=~200 spectra caused by
different C/O ratios, nitrogen abundances and masses. This is done by normalizing
them with a calculation where the investigated parameters have solar values.

\subsubsection{C/O Ratio \& Nitrogen Abundance}
\label{sec_spec1}

\begin{figure}
\includegraphics[width=9.5cm,clip]{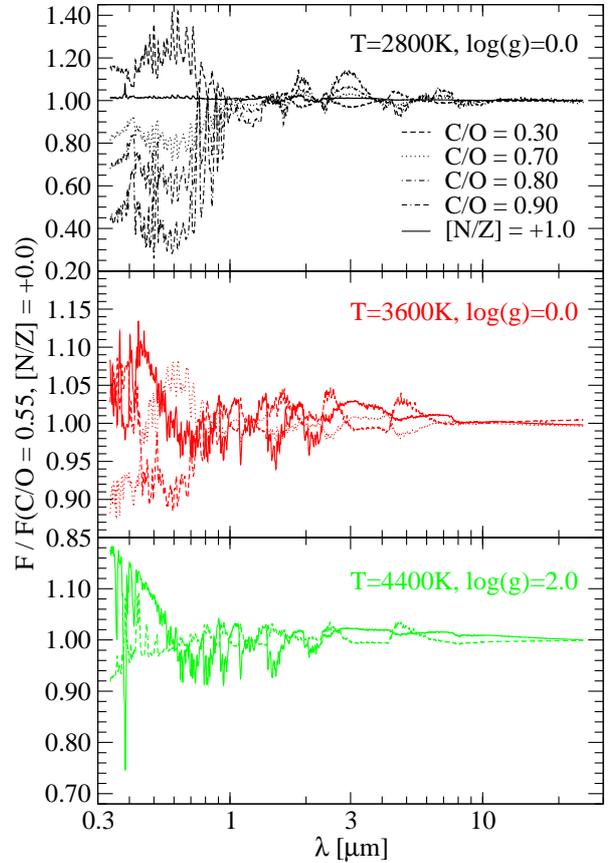}
\caption{Relative spectral changes due to different C/O ratios
and an increased nitrogen abundance for COMARCS models
with solar mass and metallicity. The spectra are normalized
to a calculation with solar composition (C/O~=~0.55,
$\rm log(\varepsilon_N/\varepsilon_H) + 12 = 7.90$).
Three examples are included: $\rm T_{eff} = 2800~K$,
$\rm log(g~[cm/s^2]) = 0.0$ (upper panel) with
C/O~=~0.30, 0.70, 0.80, 0.90, [N/Z]~=~+1.0,
$\rm T_{eff} = 3600~K$, $\rm log(g~[cm/s^2]) = 0.0$
(middle panel) with C/O~=~0.30, 0.70, [N/Z]~=~+1.0 and
$\rm T_{eff} = 4400~K$, $\rm log(g~[cm/s^2]) = 2.0$
(lower panel) with C/O~=~0.30, [N/Z]~=~+1.0.}
\label{ari_aburela}
\end{figure}

In Fig.~\ref{ari_aburela} one can see what happens, if the nitrogen abundance
increases by a factor of ten and the C/O value changes to 0.3, 0.7, 0.8 or 0.9.
In order to emphasize the effects, the spectra are normalized using a calculation
with solar chemical composition implying C/O~=~0.55 and [N/Z]~=~0. The figure
has three panels with different combinations of temperature and surface gravity.
The corresponding sets of COMARCS models are equal to the ones described in
Section~\ref{sec_atstru} and appearing in Figs.~\ref{ari_costruc} and \ref{ari_nstruc}.
Only the computations with solar abundances are missing, since they have been
used for the normalization. Thus, they would get a constant value of one in
the plot.

For the coolest of the shown sets with $\rm T_{eff} = 2800~K$ and $\rm log(g~[cm/s^2]) = 0.0$
there exists a clear trend. The absorption at short wavelengths dominated by TiO and atoms
increases with higher C/O values, while the water bands above 1.8~$\mu$m become weaker.
Considerable differences appear especially in the optical range. The flux in the
V filter varies by a factor of about three, if C/O grows from 0.3 to 0.9. But even
a moderate change from solar composition to C/O~=~0.3 or 0.7 causes deviations of
$\pm 20$\%. The relation between atmospheric structure, chemical equilibrium and
molecular absorption is often not simple. As it has already been mentioned in
Section~\ref{sec_atstru}, the abundance of TiO, which produces a strong heating,
decreases in the inner regions of the models at higher C/O ratios, while it grows
in the outer areas below 2400~K\@.

An enhancement of nitrogen has almost no effect on the spectra of the coolest
set. As one can see in Fig.~\ref{ari_molec01}, this is mainly due to the weakness
of the CN bands appearing at lower temperatures. Also the large opacities of TiO and
water in the outer regions of the cold atmospheres play a role. In K and M giants CN
is the most important molecule contributing to the absorption, which includes
nitrogen. The situation becomes quite different in the warmer sets with 3600 and
4400~K, where [N/Z] has some influence. In these cases an increased value causes
deeper CN bands and weaker atomic features. In general the variations produced by the
investigated changes of the composition are smaller at the higher temperatures. Apart
from a few exceptions they never exceed $\pm 10$\% in the two warmer sets. However, one
should not forget that at 3600 and 4400~K the C/O ratios above 0.7 are not covered.

Concerning the behaviour as a function of the C/O value there is a clear trend in the
spectra of the warmer sets. The depth of the CO bands in the near and mid infrared
increases with a higher ratio, which can be explained by the fact that more carbon
becomes available to form the molecule. This effect appears already at 2800~K, where
it is less pronounced due to the contamination with the opacity of water. In the models
with 3600~K the absorption around 0.6~$\mu$m, which is dominated by TiO, gets weaker,
if the C/O value grows. This is the opposite of the behaviour observed for the coolest
set and an example for a more complex relation, where trends may be reversed when
changing the basic stellar parameters. Thus, one should always be careful, if abundance
corrections to synthetic spectra or photometry determined for a certain combination
of effective temperature and surface gravity are transferred to models with different
properties. At 2800 and 3600~K the intensity of the atomic absorption below 0.5~$\mu$m
increases with a higher C/O ratio. Also this trend does not appear in all places, as
one can see looking at the 4400~K set in the plot.

The significance of the excess of oxygen compared to carbon (O$-$C) for the spectra
and atmospheric structures at a changing C/O ratio is discussed in the second part of the
appendix. Especially in cooler models this parameter will be important, since most of the
C atoms form CO molecules.

\begin{figure}
\includegraphics[width=9.5cm,clip]{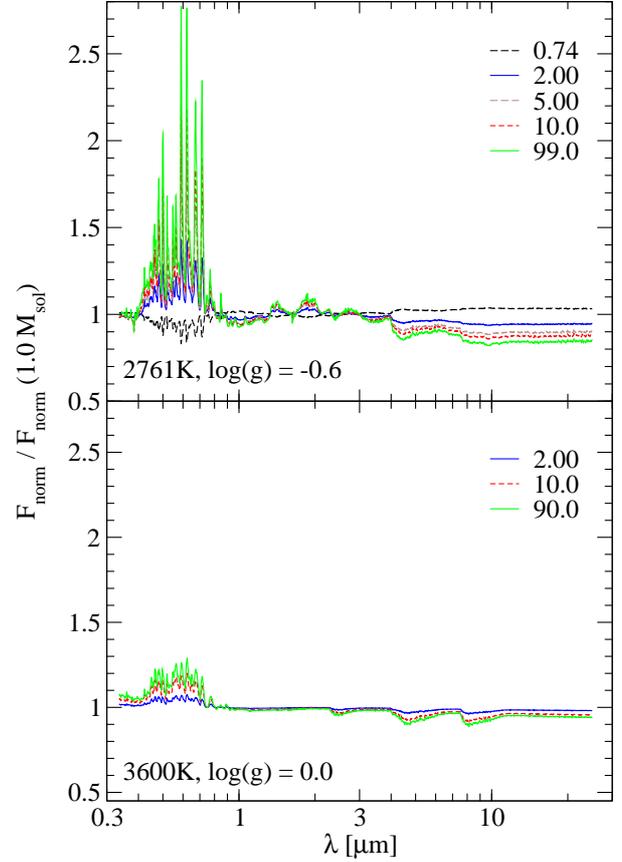}
\caption{Relative spectral changes due to different stellar mass
for COMARCS models with solar abundances. The spectra are normalized
to a calculation with $\rm M/M_{\odot} = 1.0$. The effects are shown for
parameters typical for a late AGB object ($\rm T_{eff} = 2761~K$,
$\rm log(g~[cm/s^2]) = -0.6$, upper panel) and a RSG or RGB star
($\rm T_{eff} = 3600~K$, $\rm log(g~[cm/s^2]) = 0.0$, lower panel).}
\label{ari_masssp}
\end{figure}

\subsubsection{Mass}
\label{sec_mass1}

In Fig.~\ref{ari_masssp} we investigate the impact of the stellar mass on our
COMARCS spectra. If the effective temperature and surface gravity are fixed, this
parameter represents the degree of sphericity in a hydrostatic atmosphere. It causes
changes, which are in general much smaller than the ones connected to the two other
mentioned quantities. Thus, the corresponding deviations can also be treated as a
correction applied to the result of an interpolation in the main variables, when
using the COMARCS grid to determine observable properties of stars (see
Section~\ref{sec_evo}). In Fig.~\ref{ari_masssp} we show spectra, which are
normalized relative to a calculation with one solar mass. This helps to
identify the changes caused by the investigated parameter. Two sets of models
with the chemical composition of the Sun are presented, which are characterized
by their effective temperature and surface gravity. The first one was taken from
the stellar evolution track described in Section~\ref{sec_par} and displayed in
Fig.~\ref{ari_grid}. With $\rm T_{eff} = 2761~K$ and $\rm log(g~[cm/s^2]) = -0.6$
it is typical for a very extended late AGB star. Masses between 0.74 and
99~M$_{\odot}$ are covered. The smallest value corresponds to the result from the
evolutionary track, if the loss of material due to stellar winds is considered.
It should be noted that in the used PARSEC and COLIBRI models this effect
plays only a significant role during the last stages of the AGB phase. The parameters
of the second set having $\rm T_{eff} = 3600~K$ and $\rm log(g~[cm/s^2]) = 0.0$ are
close to the regions, where one can find RSG or RGB stars, with the latter being
typically a bit more compact. In this case masses between 2 and 90~M$_{\odot}$ are
shown. The very high values were chosen to simulate a plane-parallel environment.

For both sets shown in Fig.~\ref{ari_masssp} we observe the same trend. The intensity
of the TiO bands in the optical range increases with lower mass, while the molecular
absorption in the mid and far infrared becomes smaller. This weakening affects in the
cooler models mainly the water opacity and in the warmer ones the features of CO and
SiO\@. It can be explained by a geometric phenomenon. A lower mass causes a larger
contribution of the optically thin edges of a star to the total flux, which produces
more intense emission components in the lines. These may be directly observed in the
profiles of different transitions especially in the infrared, if one studies high
resolution spectra of pulsating giants with a radial and temporal variation of the
velocity fields \citep[e.g.][]{2010A&A...514A..35N,2001A&A...379..305A}. Due to an
increased atmospheric extension the effect will become very pronounced in dynamical
objects with significant stellar winds. It is usually stronger at the longer
wavelengths, where one might in some cases even get broad emission features
\citep[e.g.][]{2002A&A...383..972M,2004A&A...422..289G}. In our hydrostatic
COMARCS models with no mass loss or velocity fields we obtain only the mentioned
weakening of the lines.

The changes of the TiO bands visible in Fig.~\ref{ari_masssp} are caused by
differences of the atmospheric structure. As one would expect, all effects of mass
are much larger for the more extended cooler set, since they depend a lot on the
surface gravity. If $\rm log(g~[cm/s^2])$ grows above zero, they become quickly
very weak. Thus, the sphericity needs only to be taken into account during the
last (and maybe first) stages of stellar evolution. However, such stars are often
far away from hydrostatic equilibrium.

\subsection{Synthetic Photometry}
\label{sec_phot}

\begin{figure}
\includegraphics[width=7.0cm,clip,angle=270]{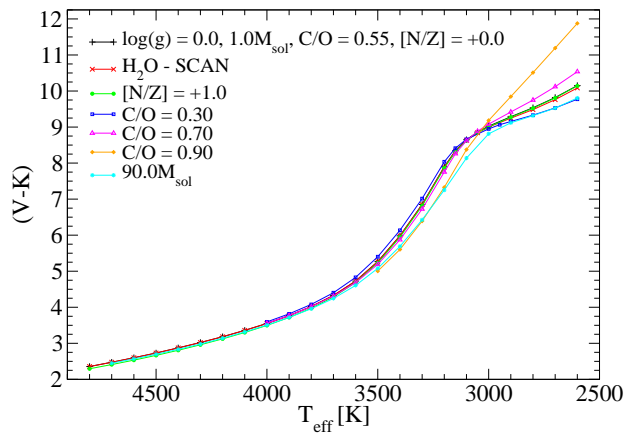}
\caption{(V$-$K) as a function of the effective temperature for COMARCS
models with solar metallicity and $\rm log(g~[cm/s^2]) = 0.0$. The standard
parameters are $\rm M/M_{\odot} = 1.0$, solar abundances (C/O~=~0.55, [N/Z]~=~0.0)
and usage of the BT2 linelist for water below 4000~K (black). The result of changing
C/O to 0.3 (blue), 0.7 (magenta) and 0.9 (orange), of increasing nitrogen by a
factor of ten ([N/Z]~=~+1.0, green) or the mass to 90~M$_{\odot}$ (cyan) is shown.
In addition, we demonstrate the effect of taking the H$_2$O opacities from the
SCAN data (red).}
\label{ari_tvk1}
\end{figure}

The synthetic photometry computed for the complete grid of our hydrostatic COMARCS
models can be found in \url{http://starkey.astro.unipd.it/atm}. The tables
contain the bolometric corrections (BC) described in Section~\ref{sec_syn} as a function
of the stellar properties. The corresponding main parameters are $\rm T_{eff}$,
$\rm log(g~[cm/s^2])$, $\rm M/M_{\odot}$, [Fe/H], [O/H], [C/H], $\xi$~[km/s] and a
flag for the used water absorption data (SCAN or BT2, see Section~\ref{sec_opac}).
[Fe/H] which represents [Z/H] as well as [O/H] and [C/H] which determine the C/O
ratio are listed with the absolute abundances scaled relative to hydrogen
$\rm log(\varepsilon_{Fe,O,C}/\varepsilon_H) + 12$. Other quantities like [N/H], [Y/H]
or [Zr/H] were included, if they deviate from the standard values. The opacity sets
available at the moment in the COMARCS grid can be found in Table~\ref{ari_comarcs}.

The complete set of tables covers the bolometric corrections for more than 40 photometric
systems, which are separated into different files containing the values of all included
filters and COMARCS models. This wealth of information may be exploited by the users of
the database. In the current work we will only discuss the visual and near infrared
magnitudes that were determined according to the definition from \citet{1990PASP..102.1181B}
and \citet{1988PASP..100.1134B}. The V, I, J, H and K values and corresponding colours,
which appear in the following sections and figures, are based on this photometric system
(Bessell). The mentioned filters have been used for a large number of observations.

The observed colours of many cool and variable stars are severely affected by reddening
due to circumstellar dust \citep{2011A&A...529A.129N}. Since hydrostatic model atmospheres
do not take this process into account, the corresponding corrections have to be applied
a posteriori to the results \citep{2008A&A...482..883M}. However, in such cases it is
preferable to use dynamical calculations describing pulsation and mass loss.

\begin{figure}
\includegraphics[width=7.0cm,clip,angle=270]{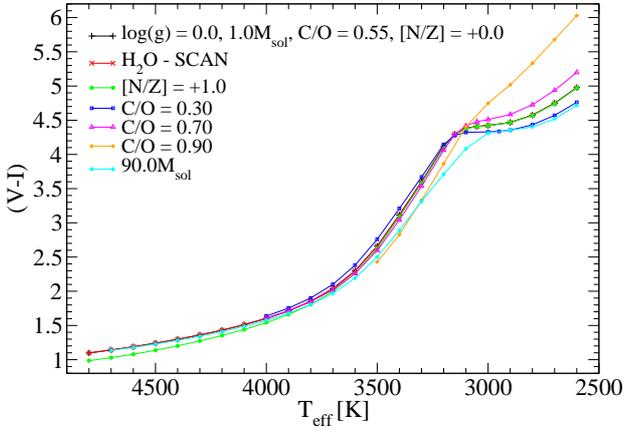}
\caption{(V$-$I) as a function of the effective temperature for COMARCS
models with solar metallicity and $\rm log(g~[cm/s^2]) = 0.0$. Parameters,
symbols and lines are like in Fig.~\ref{ari_tvk1}.}
\label{ari_tvi1}
\end{figure}

\begin{figure}
\includegraphics[width=7.0cm,clip,angle=270]{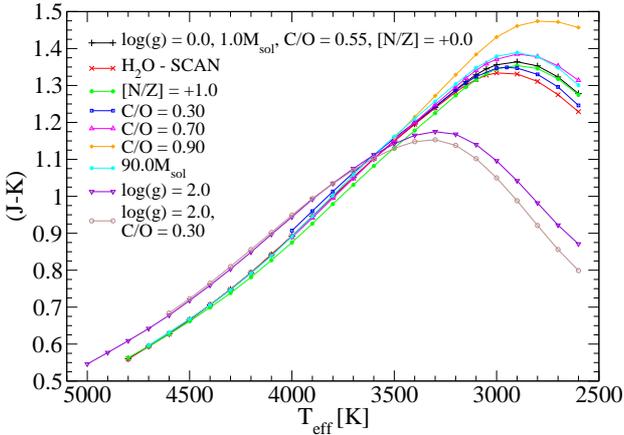}
\caption{(J$-$K) as a function of the effective temperature for COMARCS models
with solar metallicity. The standard parameters are $\rm log(g~[cm/s^2]) = 0.0$,
$\rm M/M_{\odot} = 1.0$, solar abundances (C/O~=~0.55, [N/Z]~=~0.0) and usage of
the BT2 linelist for water below 4000~K (black). The result of changing
C/O to 0.3 (blue), 0.7 (magenta) and 0.9 (orange), of increasing nitrogen by a
factor of ten ([N/Z]~=~+1.0, green) or the mass to 90~M$_{\odot}$ (cyan) is shown.
In addition, we demonstrate the effect of taking the H$_2$O opacities from the
SCAN data (red) and of adopting $\rm log(g~[cm/s^2]) = 2.0$ with C/O~=~0.55
(violet) or 0.3 (brown).}
\label{ari_tjk1}
\end{figure}

\begin{figure}
\includegraphics[width=7.0cm,clip,angle=270]{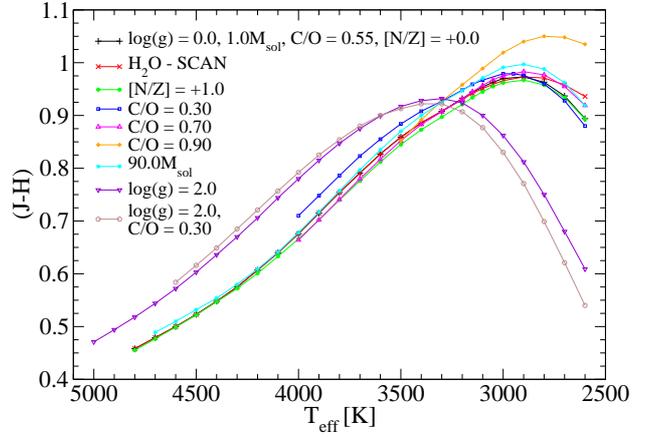}
\caption{(J$-$H) as a function of the effective temperature for COMARCS
models with solar metallicity. Parameters, symbols and lines are like
in Fig.~\ref{ari_tjk1}.}
\label{ari_tjh1}
\end{figure}

In Figs.~\ref{ari_tvk1}, \ref{ari_tvi1}, \ref{ari_tjk1} and \ref{ari_tjh1} the colours
(V$-$K), (V$-$I), (J$-$K) and (J$-$H) are presented as a function of the effective
temperature for COMARCS atmospheres with solar metallicity. We compare results obtained
with a C/O ratio of 0.3, 0.7 and 0.9, [N/Z]~=~+1.0, 90~M$_{\odot}$ as well as
$\rm log(g~[cm/s^2]) = 2.0$ to a standard model having $\rm log(g~[cm/s^2]) = 0.0$,
1.0~M$_{\odot}$ and the chemical abundances of the Sun. In addition, we show the
consequences of using the SCAN water data instead of BT2 for stars cooler than
4000~K (see Section~\ref{sec_opac}). Calculations with $\rm log(g~[cm/s^2]) = 2.0$
are only included in the plots for (J$-$K) and (J$-$H), since these colours are
more affected by the surface gravity.

\subsubsection{C/O Ratio}

As one can see in Figs.~\ref{ari_tvk1} and \ref{ari_tvi1}, (V$-$K) and (V$-$I) show a
quite similar behaviour, if the value of C/O is changed. Below 3000 to 3100~K the colours
at higher ratios become significantly redder. For the coolest models and C/O~=~0.9 the
deviations grow up to +2~mag in (V$-$K) and +1~mag in (V$-$I). When the effective
temperatures exceed the mentioned limit, the trend reverses. We obtain bluer colours
for an increasing C/O ratio, although the changes always remain smaller than in the
colder giants. Even at C/O~=~0.9 the absolute values stay below about 0.5~mag in
(V$-$K) and 0.3~mag in (V$-$I). The behaviour observed here is mainly caused by a
variation of the TiO absorption in the V filter, which has been discussed in
Section~\ref{sec_spec1}. In the cooler models the intensity of these molecular bands
increases with a higher C/O ratio creating redder colours, while for warmer temperatures
the opposite trend may appear (see Fig.~\ref{ari_aburela}).

The (J$-$K) colour shown in Fig.~\ref{ari_tjk1} also gets redder, if the C/O value is
increased in cooler objects. This phenomenon appears at effective temperatures below
about 3400~K and for both surface gravities included in the plot. At C/O~=~0.9 the
deviations may become quite large exceeding +0.15~mag in the coldest giants. They are
mainly due to a weakening of the water bands above 1.8~$\mu$m with higher C/O ratios
as it is discussed in Section~\ref{sec_spec1} and visible in Fig.~\ref{ari_aburela},
since this produces more flux in the K filter. In the models warmer than 3400~K changes
of the C/O value cause only very small variations of the (J$-$K) colour.

The relation between (J$-$H) and C/O is a bit more complex. Like for (J$-$K) the colour
index of the coolest models increases with the ratio. As one can see in Fig.~\ref{ari_tjh1}
this clear trend disappears for the objects with $\rm log(g~[cm/s^2]) = 0.0$ and a low
C/O value already at 2900~K, while it continues up to 3200~K for the compact and
carbon-rich sequences. At higher effective temperatures we even observe a reversion
of the behaviour: the stars with C/O~=~0.3 are predicted to be significantly redder than those with
a larger ratio. This phenomenon is especially pronounced in the more extended models,
where it appears above 3300~K\@. The increase of the colour index may exceed +0.03~mag
and correspond almost to a 100~K shift of the effective temperature. The studied changes
of (J$-$H) are mostly due to variations of the CO bands and in cooler objects of the
water absorption.

\subsubsection{Nitrogen Abundance}

Among the four colour indices investigated here only (V$-$I) is significantly changed by an
enhancement of nitrogen. Models warmer than 4000~K get bluer, if the abundance of that
element increases. As one can see in Fig.~\ref{ari_tvi1} the shift for [N/Z]~=~+1.0 may
exceed 0.1~mag, which corresponds to a difference of the effective temperature growing
up to about between 100 and 200~K in this range. Looking at Fig.~\ref{ari_aburela} it
becomes clear that the deviations of (V$-$I) are due to a combination of weaker atomic
absorption in the V and deeper CN features in the I filter, if nitrogen is more enhanced.
Below 4000~K the changes disappear completely.

An increase of the CN band intensity with a higher nitrogen abundance may also produce
small variations of the (J$-$K) and (J$-$H) colours. It is shown in Figs.~\ref{ari_tjk1}
and \ref{ari_tjh1} that the models in the range between 3200 and 4200~K get bluer for
[N/Z]~=~+1.0. However, these changes are not significant, since they always remain
clearly below the level equivalent to a 50~K difference of the effective temperature.

\subsubsection{Mass}

In order to study the effects of sphericity we compare in Figs.~\ref{ari_tvk1},
\ref{ari_tvi1}, \ref{ari_tjk1} and \ref{ari_tjh1} models calculated with 90~M$_{\odot}$
to the ones in the standard sequence having the mass of the Sun. The high value was
chosen, since it produces results close to a plane-parallel solution. The corresponding
colour shifts also represent in most cases an upper limit for the changes expected at a
certain combination of effective temperature, surface gravity and chemical abundances.
One can see in Fig.~\ref{ari_masssp} that the variation of the spectra as a function
of the mass is monotonous and regular. As it has been mentioned in Section~\ref{sec_mass1}
the deviations caused by sphericity increase a lot, when the value of log~(g) becomes
lower. Above $\rm log(g~[cm/s^2]) = 0.0$ they disappear very soon.

The colour indices (V$-$K) and (V$-$I) presented in Figs.~\ref{ari_tvk1} and
\ref{ari_tvi1} show a very similar behaviour. Below 3700 to 3800~K the models become
bluer, when the mass increases. The maximum shift between 1.0 and 90~M$_{\odot}$, which
appears around 3200~K, exceeds 0.5~mag in (V$-$K) and 0.4~mag in (V$-$I). As it has
been noted before these deviations will grow at lower surface gravity. The bluer colours
are mainly caused by the weakening of the TiO bands in the V filter, which is visible
in Fig.~\ref{ari_masssp}.

One can see in Figs.~\ref{ari_tjk1} and \ref{ari_tjh1} that the changes of (J$-$K) and
(J$-$H) due to sphericity effects also resemble each other. Below about 3500 to 3600~K
the colours get redder as the mass increases. In the cooler stars this is at least
partly caused by a variation of the water absorption. However, for the surface gravity
used in the plots the deviations never become very pronounced.

\subsubsection{Water Opacity and General Remarks}

As one would expect (V$-$I) is not and (V$-$K) almost not affected by the selection
of the water list, since there exists no significant absorption of this molecule
below 1.0~$\mu$m, while the changes in the K filter visible in Fig.~\ref{ari_lists2}
are small compared to the colour shifts due to a different temperature. For (J$-$K)
and (J$-$H) we find some moderate deviations appearing in the coolest models, which
can in both cases reach values of about 0.05~mag. When the SCAN data are used, (J$-$K)
becomes bluer and (J$-$H) redder. For the giants investigated here the changes occur
only below 3000 to 3100~K\@. At this point we want to note that because of uncertainties
concerning the stellar parameters and possible deviations from hydrostatic structures
with spherical symmetry as well as from LTE, which appear in cool stars, it is not
easy to select the preferable line list by studying the available infrared observations.

Compared to the effects of temperature, surface gravity and metallicity the variations
due to changes of the C/O ratio, nitrogen abundance and mass are usually smaller. However,
in many regions of the HRD they are not negligible. Thus, they need to be included at least
as correction factors when deriving synthetic spectra and photometry from grids of models.
An exception are C/O values close to one in cool stars, where large deviations of the
observable properties may appear. But in those objects also the enhancement of s-process
elements has to be considered, which will mainly result in much deeper ZrO bands. As we
have already mentioned before, this effect is not covered in the current work. More
information on models of S-type giants can be found in \citet{2011ASPC..445...71V}.

\section{Discussion}
\label{sec_discussion}

In the following text we investigate the photometric results obtained from the COMARCS
grid, which are described in Section~\ref{sec_phot}, regarding observations and the
application in connection with stellar evolution calculations. Three aspects will be
discussed. First, we present a comparison to empirical colour-colour and semi-empirical
colour-temperature relations from the literature. Second, the observed and simulated
photometric properties of stars in the Galactic Bulge are studied. Finally, we explore
the general use of the database in association with results from synthetic stellar
evolution to describe larger samples of objects.

\subsection{Comparison with Observed Relations}

Using a large collection of observational material available in the literature
\citet{2011ApJS..193....1W} have produced a database relating different colours
with each other and with the main stellar parameters. We have applied their tables
in combination with the connected interpolation routines to calculate the photometric
properties of the stars situated on the evolutionary track for solar mass and abundances,
which is described in Section~\ref{sec_par} and shown in Fig.~\ref{ari_grid}. The
results can be compared to the ones obtained from our COMARCS models. Especially the
colour-colour relations are interesting, since they were directly taken from the
observations with a minimum of additional assumptions. One of these conditions is
that the chemical abundances are known and correspond to a solar mixture scaled
with the metallicity.

\begin{figure}
\includegraphics[width=8.0cm,clip,angle=0]{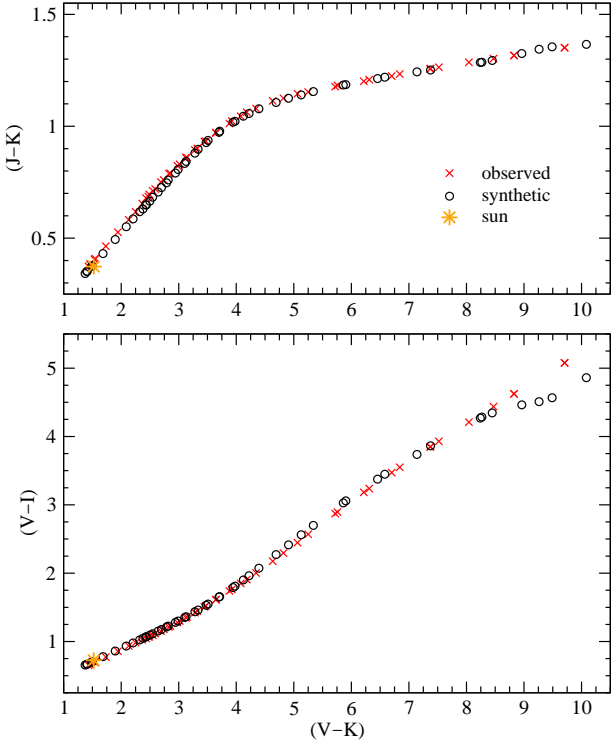}
\caption{Two colour diagrams (J$-$K) and (V$-$I) versus (V$-$K) for
stars placed along an evolutionary track with one solar mass and
$\rm Z/Z_{\odot} = 1.0$ (see Fig.~\ref{ari_grid}). The results for a sequence
of COMARCS models (black circles) are compared to the corresponding values
taken from the fit routines by \citet{2011ApJS..193....1W}, which are based
on a large collection of observational material (red crosses). The colours
of the Sun according to \citet{1998A&A...333..231B} are also shown (yellow
stars).}
\label{ari_worthey_vk}
\end{figure}

In Fig.~\ref{ari_worthey_vk} we show (J$-$K) and (V$-$I) as a function of
(V$-$K), which is in general considered to be a good indicator for the effective
temperature. In addition to the results obtained from the COMARCS atmospheres
and the routine from \citet{2011ApJS..193....1W} we have also included the position
of the Sun according to \citet{1998A&A...333..231B}. The agreement between the
observed and synthetic colour-colour relations presented in the plots is in most
of the cases quite satisfactory. Significant deviations appear only in (V$-$I) for
stars cooler than about 3200~K, where our models are bluer at constant (V$-$K).
However, for such cold giants one should expect differences. Most of them show
strong variations, which is a problem, if no simultaneous photometric measurements
are available. Since the connected pulsations also create considerable deviations
from hydrostatic structures, the predictions based on the COMARCS atmospheres
and the existing abundance determinations can often not be trusted
\citep[e.g.][]{2014A&A...567A.143L}. In addition, many of the objects are
reddened by circumstellar dust, and the effects of mass and a changed C/O ratio
discussed in Section~\ref{sec_phot} and modifying the TiO band intensity have to be
taken into account.

\begin{figure}
\includegraphics[width=8.0cm,clip,angle=0]{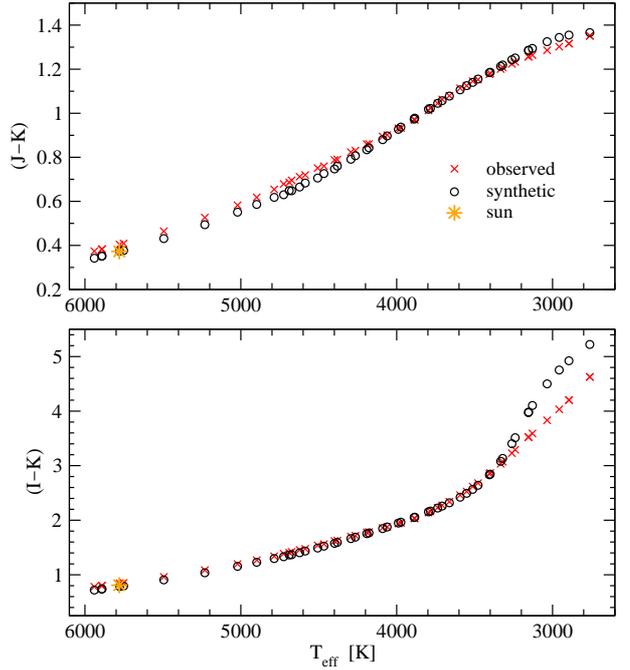}
\caption{(J$-$K) and (I$-$K) as a function of the effective temperature
for stars placed along an evolutionary track with one solar mass and
$\rm Z/Z_{\odot} = 1.0$ (see Fig.~\ref{ari_grid}). The presented data
and the plot symbols are the same as in Fig.~\ref{ari_worthey_vk}}.
\label{ari_worthey_teff}
\end{figure}

In Fig.~\ref{ari_worthey_teff} we show the (J$-$K) and (I$-$K) colours as a function
of the effective temperature. The predictions taken from our COMARCS models are again
compared to values based on observations. The corresponding relations from
\citet{2011ApJS..193....1W} have been derived using a number of sources involving
various theoretical and semi-empirical approaches. Especially for the coolest giants
the determination of the effective temperature is in general accompanied by considerable
uncertainties. In addition to the issues mentioned in the previous paragraph there are
also problems concerning the measurement of the distances and radii of such extended
objects. Thus, the differences between predicted and observed (I$-$K) indices below
3300~K appearing in the second panel of the diagram are expected, while we find a quite
good agreement of this colour for all warmer stars. The same applies to (V$-$K), which
is not shown here, since it behaves in a very similar way. We note that due to the
divergent (V$-$K) values of the coolest giants the spacing of points in
Fig.~\ref{ari_worthey_vk} is not equal at the low temperatures.

The predicted and observed (J$-$K) indices in the upper panel of Fig.~\ref{ari_worthey_teff}
agree in general quite well. However, the COMARCS models give above 4200~K slightly bluer
and below 3200~K redder colours than the relations from \citet{2011ApJS..193....1W}. This
effect occurs also in the (J$-$K) versus (V$-$K) diagram in Fig.~\ref{ari_worthey_vk},
where it is less pronounced. The solar values from \citet{1998A&A...333..231B}, which
are shown in the plot, can be found very close to our predicted sequence. It is possible
that the shifts may be explained by small differences in the definition of the photometric
filters, although \citet{2011ApJS..193....1W} use the same system in the near infrared.
Changes caused by a decreased microturbulent velocity are not sufficient to obtain the
deviations.

The (J$-$K) values as a function of effective temperature predicted by
\citet{1998A&A...333..231B} in the range from 3600 to 5000~K for stars with
$\rm log(g~[cm/s^2]) = 0.0$ and 2.0 are very similar to our COMARCS results with
deviations below 0.025~mag in most of the cases. Somewhat larger differences
appear only for the coolest models in the interval. In \citet{2011ApJS..193....1W} one can
find plots where they compare a few of their relations concerning colours and bolometric
corrections, which have been used here, to published theoretical work.

\subsection{The Galactic Bulge}

\begin{figure*}
\includegraphics[width=11.0cm,height=20.0cm,clip,angle=270]{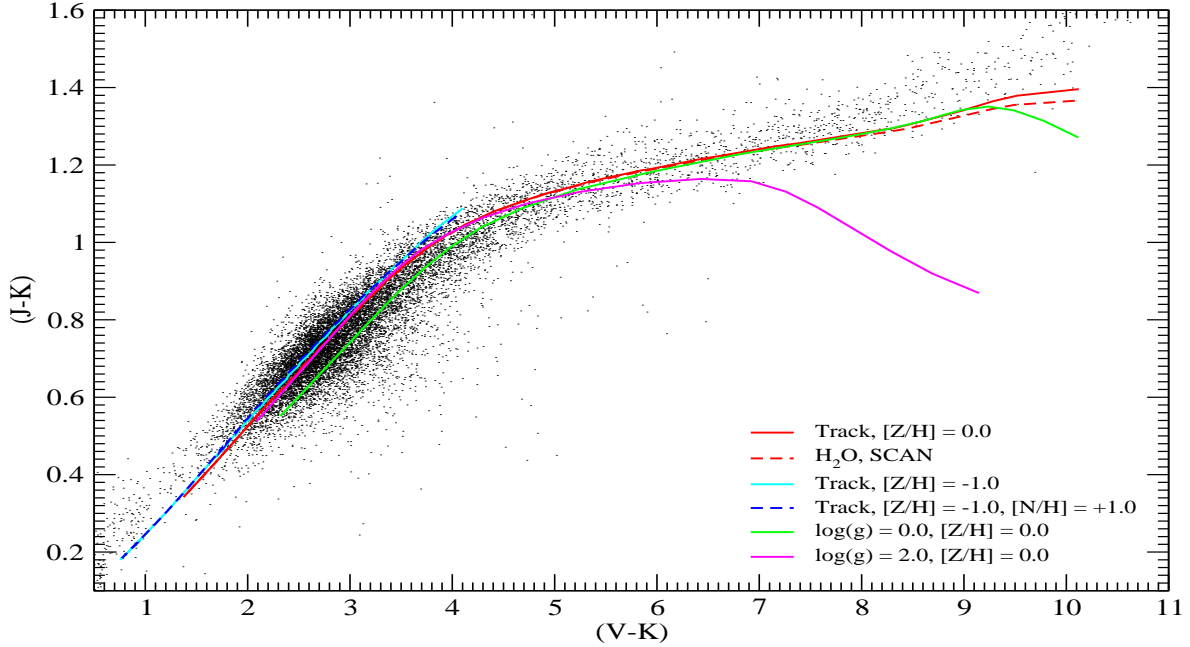}
\caption{Two colour diagram (J$-$K) versus (V$-$K) for different sequences of
COMARCS atmospheres. Models placed along evolutionary tracks for stars with
one solar mass and $\rm Z/Z_{\odot} = 1.0$ (full line, red) as well as
$\rm Z/Z_{\odot} = 0.1$ (full line, cyan) are displayed (see Fig.~\ref{ari_grid}).
For the solar metallicity track we show the effect of using the SCAN data for
water (dashed line, red) and for the metal-poor one the result of increasing
nitrogen by a factor of ten ([N/Z]~=~+1, dashed line, blue). In addition, two sequences
of models having solar mass and abundances with $\rm log(g~[cm/s^2]) = 0.0$ (full line,
green, $\rm T_{eff}$ 2600 to 4800~K) and $\rm log(g~[cm/s^2]) = 2.0$ (full line, magenta,
$\rm T_{eff}$ 2600 to 5000~K) are included. The dots represent cool giants in Baade's
window (black) and Mira variables (orange) in the Galactic Bulge.}
\label{ari_bulgvkjk}
\end{figure*}

\begin{figure*}
\includegraphics[width=17.5cm,height=10.5cm,clip,angle=0]{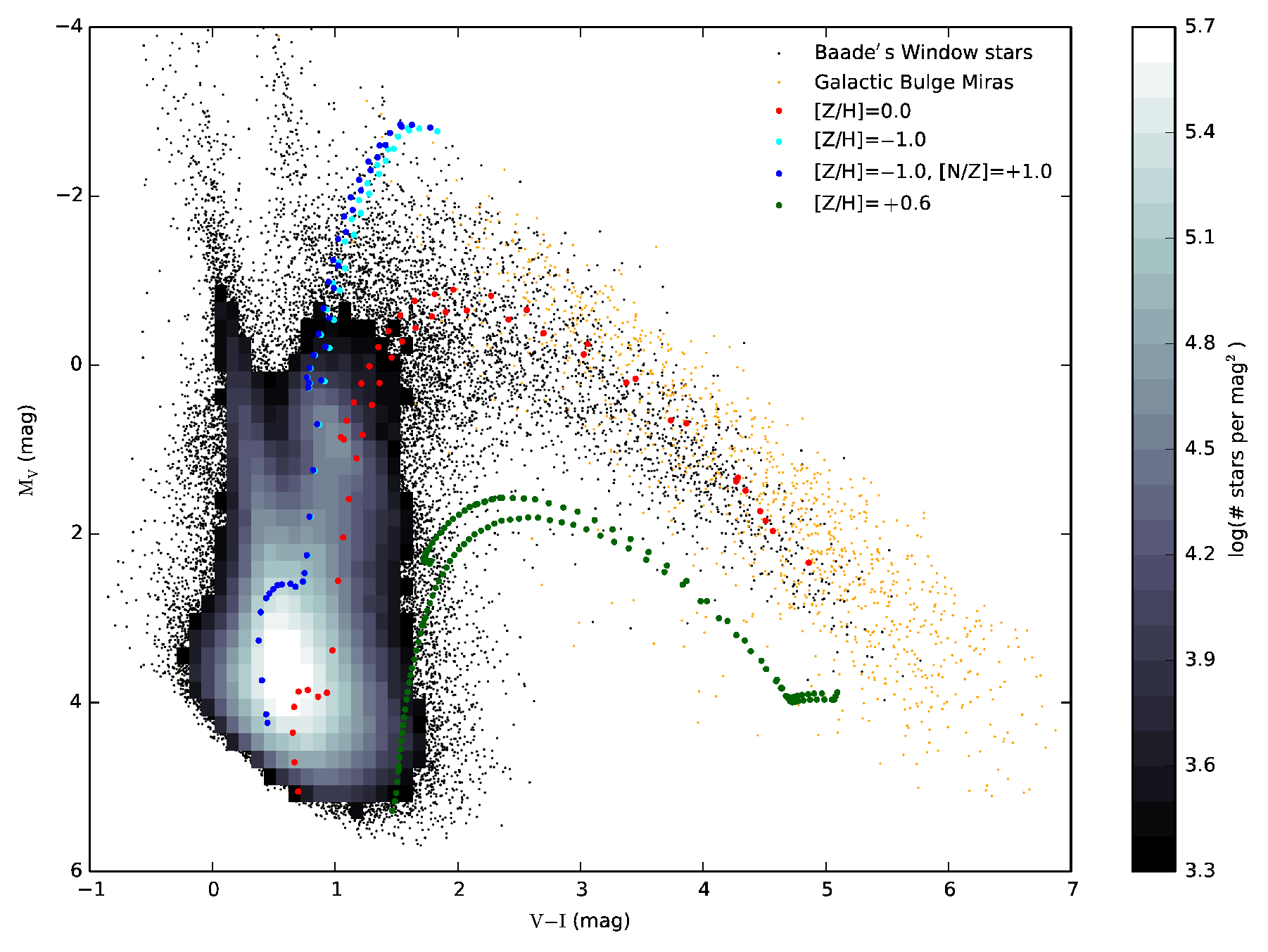}
\caption{Colour-magnitude diagram $\rm M_V$ versus (V$-$I) for COMARCS models
placed along evolutionary tracks for stars with one solar mass and
$\rm Z/Z_{\odot} = 1.0$ (red circles) as well as $\rm Z/Z_{\odot} = 0.1$
(cyan circles). For the metal-poor track we show the effect of increasing
nitrogen by a factor of ten ([N/Z]~=~+1, blue circles). The dots represent
cool giants in Baade's window (black) and Mira variables (orange) in the
Galactic Bulge. Due to the large number of objects in the first group densities
are plotted in the central region of their distribution. In addition, we
show the points of a metal-rich evolutionary track with one solar mass and
$\rm Z/Z_{\odot} = 3.98$ (green circles), which were obtained from an interpolation
in the COMARCS grid.}
\label{ari_bulgvimv}
\end{figure*}

\subsubsection{Observational Data}

Simultaneous broad-band photometric measurements in the visual and the near infrared for a whole
population are only sparsely available in the literature. Therefore we compiled data covering
the Galactic Bulge from different sources. Being dominated by cool giants located in the inner
regions of the Milky Way these observations can be compared to the modelling results as shown in
Figs.~\ref{ari_bulgvkjk} and \ref{ari_bulgvimv}.

First, we adopted from public-domain surveys photometric observations in the direction of
Baade's window, which is one of the low extinction windows towards the centre of our galaxy
\citep[e.g.][]{2002A&A...381..219D}. For the visual filters V and I we made use of the
photometry obtained by the OGLE-II program and downloaded the data of the Galactic Bulge
fields \texttt{bul\_sc1} and \texttt{bul\_sc45}, which cover Baade's window
\citep[cf.][]{2002AcA....52..217U,2004MNRAS.349..193S}. Based on the exhaustive photometric
time series of OGLE in I and an additional less extensive one in V, \citet{2002AcA....52..217U}
provide mean magnitudes for all objects with values tied to the \citet{1992AJ....104..372L}
standards. Hence, no conversion to the Bessell system had to be applied. Dereddening was done
with the help of extinction data derived by \citet{2004MNRAS.349..193S} individually for the
OGLE-II fields. The resulting colour-magnitude diagram for this rich population of about
530000 stars can be seen in Fig.~\ref{ari_bulgvimv}. The visual data were subsequently
cross-correlated with near infrared photometry of the 2MASS All-Sky Point Source Catalog
\citep{2003yCat.2246....0C}, which is made public by the NASA/IPAC Infrared Science
Archive (IRSA). Corresponding roughly to the above mentioned OGLE fields in the Galactic
Bulge, we extracted photometric J, H and $\rm K_S$ measurements for an area of
$\Delta \alpha \times \Delta \delta$~$\approx$~0.5$^{\circ} \times$~1$^{\circ}$. The correction
for interstellar reddening was carried out according to \citet{2002A&A...381..219D} and
\citet{1990ARA&A..28...37M}. The 2MASS data had to be transformed to the Bessell system, chosen
for the comparison in this work, using the relations provided by \citet{2001AJ....121.2851C}.
As the 2MASS survey reaches not as deep as OGLE, the cross-correlation results in significantly
less objects (about 14000), which are shown in Fig.~\ref{ari_bulgvkjk}.

Second, we used for the comparison with our COMARCS models a sample of M-type Mira variables
identified by \citet{2005A&A...443..143G}, who investigated the photometric variations appearing
in the OGLE-II survey. These targets are not only located in Baade's window, but spread across
the whole Galactic Bulge. \citet{2005A&A...443..143G} cross-correlated their list of objects
with the 2MASS catalogue to get single epoch J, H and $\rm K_S$ photometry for the stars, which
we will use here. The data have been dereddened adopting the interstellar extinction values
for the various OGLE-II fields as derived by \citet{2005MNRAS.364..117M}. For the
conversion to magnitudes in the Bessell system we applied again the equations of
\citet{2001AJ....121.2851C}. In addition to the near infrared photometry, Groenewegen
(priv.~comm.) cross-correlated the selected Mira stars with the OGLE ``photometric maps''
\citep{2002AcA....52..217U} and provided mean V and I magnitudes, which we dereddened
following \citet{2004MNRAS.349..193S}. This led to a sample of roughly 1000 objects that
are also plotted in Figs.~\ref{ari_bulgvkjk} and \ref{ari_bulgvimv}.

When necessary, the apparent magnitudes of the Bulge objects were transformed into absolute
ones assuming a distance modulus of (m~$-$~M)~=~14.7~mag for the centre of our galaxy
\citep[e.g.][]{2009A&A...498...95V}.

\subsubsection{Models and the Bulge Population}

In Fig.~\ref{ari_bulgvkjk} we present a two colour diagram (J$-$K) versus (V$-$K), where we
compare results obtained from different series of COMARCS atmospheres to measured values of
the giants and Mira variables in the Galactic Bulge. First, the plot includes two
sequences of models placed along stellar evolution tracks for one solar mass with [Z/H]~=~0
and $-1$, which are described in Section~\ref{sec_par} and displayed in Fig.~\ref{ari_grid}.
The calculations having the metallicity of the Sun were already used before, when we discussed
the relations of \citet{2011ApJS..193....1W}. They produce colours, which are almost for the
complete range covered by the diagram located in the centre of the distribution of the observed
stars. Deviations appear mainly for the coolest effective temperatures, where the (J$-$K) indices
of the Bulge objects tend to be higher. There exists also a sparsely populated red tail extending
up to (V$-$K)~=~14 and (J$-$K)~=~3, which is not visible in the plot and far away from the shown
COMARCS atmospheres. It contains mostly Mira variables. We have already noted that such cool
giants can usually not be described by simple hydrostatic models, since they are dominated by
pulsation, convection, mass loss and dust formation. In addition, the connected temporal changes
may create problems concerning the observed photometric values, if they were not taken at the
same time. Very red (V$-$K) and (J$-$K) colours can also be produced by COMARCS atmospheres with
a high C/O ratio between 0.8 and 1.0 (see Figs.~\ref{ari_tvk1} \& \ref{ari_tjk1}) or with a
combination of an increased metallicity and a quite low effective temperature. The first of
these scenarios is not so probable, since we find no classic carbon stars in the Galactic
Bulge. The good agreement of the models placed along the [Z/H]~=~0 track with the bulk
of the observed giants confirms the result from the comparison with the relations of
\citet{2011ApJS..193....1W} in Fig.~\ref{ari_worthey_vk}, where we see only weak deviations.

For the stellar evolution track with [Z/H]~=~$-1$ we show an additional sequence of COMARCS
atmospheres, where nitrogen was enhanced by a factor of ten. However, there is almost no
difference between the series with [N/Z]~=~0 and +1 in the diagram. One can see already in
Figs.~\ref{ari_tvk1} and \ref{ari_tjk1} that the impact of this parameter on (V$-$K) and
(J$-$K) remains quite small. Both low metallicity sequences cover only the blue part of the
plot. Due to the high effective temperatures of the corresponding stellar evolution track
they do not exceed values around 4.2 in (V$-$K) and 1.1 in (J$-$K).

In Fig.~\ref{ari_bulgvkjk} we have also added some extra series of COMARCS atmospheres with
solar abundances. For the cooler models located on the evolutionary track a sequence is
shown, which was obtained from spectra calculated using the SCAN line list for water as
described in Section~\ref{sec_opac}. One can see that the selection of the H$_2$O opacity
has a small influence on the position in the diagram. The corresponding changes are not
sufficient to decide, which data source is better. In order to prove the importance of
adopting the correct surface gravity, when synthetic photometry is derived from models, we
have also included two effective temperature sequences with solar mass and a constant
$\rm log(g~[cm/s^2])$ of 0.0 and +2.0. Both of them can only explain a part of the observed
distribution of giants in the Galactic Bulge.

In Fig.~\ref{ari_bulgvimv} we present a colour-magnitude diagram $\rm M_V$ versus (V$-$I),
where we compare again COMARCS atmospheres located on stellar evolution tracks for one
solar mass with [Z/H]~=~0 and $-1$ to giants in the Galactic Bulge. Also the calculations
with [N/Z]~=~+1 were included for the lower metallicity. The sequence of models having the
chemical abundances of the Sun agrees quite well with the observed stars. It covers almost
the complete colour range being always very close to the centre of their distribution. Only
the reddest objects, which were usually classified as Mira variables, are not reached by the
COMARCS atmospheres. As it has already been discussed, such giants cannot be described by
hydrostatic models. We may conclude that stars with a mass and metallicity similar to the
solar values are typical for the investigated Galactic Bulge population. Independent of the
nitrogen abundance, which causes small changes in (V$-$I), the sequence with [Z/H]~=~$-1$
covers only quite blue colours in the diagram. This is due to the high effective temperatures
of the corresponding stellar evolution tracks. As one can see the most luminous of these
models are located in a region, where the density of observed stars is very low. They appear
to be not so common in the studied Bulge population.

In Fig.~\ref{ari_bulgvimv} we have also included the points of a metal-rich evolutionary
track with one solar mass and [Z/H]~=~+0.6, which were obtained from an interpolation in the
COMARCS grid as described in the following Section~\ref{sec_evo}. The redder stars of this
sequence are obviously as well not common in the investigated region of the Bulge. Their low
fluxes in the V filter can be explained by the fact that at higher [Z/H] the models having a
given luminosity become in general cooler, while the depth of the TiO bands and many atomic
lines increases even at a constant effective temperature. Using only the data shown in the
plot, it is not possible to exclude the existence of a considerable population of evolved
giants with a certain larger or smaller metallicity, since they may have a different mass
(age) or [O/Z]. This would require a more detailed study covering a variation of the two
parameters.

\subsection{Hydrostatic Model Atmospheres and Stellar Evolution}
\label{sec_evo}

\begin{figure*}
\centering
\includegraphics[width=15.0cm,clip,angle=270]{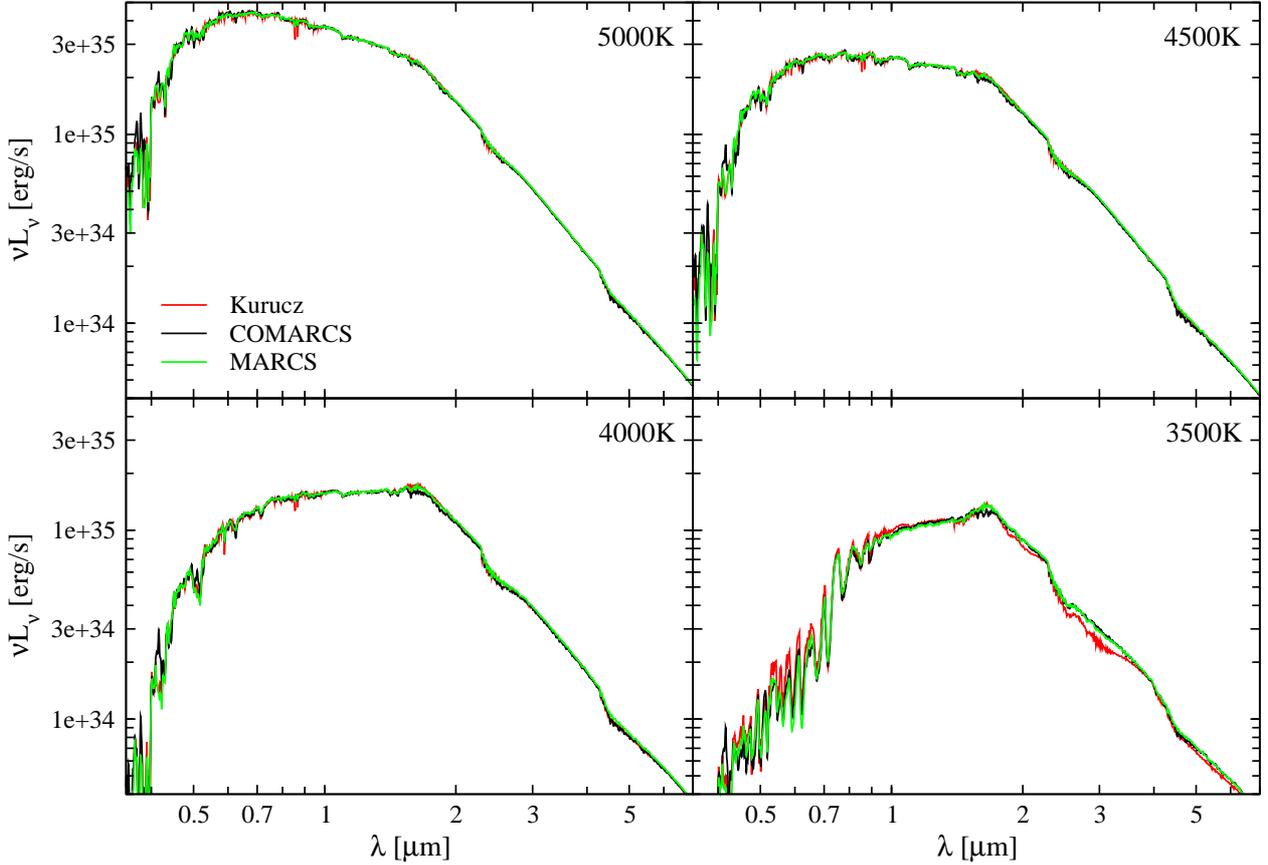}
\caption{Spectra based on COMARCS models with $\rm log(g~[cm/s^2]) = 2.0$,
$\rm M/M_{\odot} = 1.0$, solar abundances and four different effective temperatures
(5000, 4500, 4000, 3500~K, black) are compared to published MARCS (green) and
Kurucz (red) spectra corresponding to identical parameters. In the case of
COMARCS and MARCS the original OS resolution of the radiative transfer has been
degraded to R~=~200. The Kurucz spectra (ODF) are plotted without any modification
resulting in a somewhat higher resolution of narrow features. Also the solar
abundances used for the model grids differ a bit.}
\label{ari_kuru}
\end{figure*}

The bulk of the models presented in this paper can be used to define a three-dimensional
grid with $\rm T_{eff}$, log~(g) and [Z/H] as parameters. For the sake of transforming
theoretical results obtained from COMARCS atmospheres into absolute magnitudes and
colours, we make the interpolation first in the effective temperature versus surface
gravity grids of the confining metallicity values. Then we interpolate in [Z/H]. A scheme
identical to the one described in Section~4.2 of \citet{2009A&A...503..913A} was applied.

The other parameters investigated in Section~\ref{sec_results}, which are C/O ratio,
nitrogen abundance and stellar mass, have only been covered by sparser grids or even
more limited sequences of models. However, they are demonstrated not to be as crucial
as the main quantities. In order to consider their effect we simply apply corrections
to the photometric results obtained from the interpolation, which depend on temperature
and surface gravity. For each of the involved parameters this is done separately. It allows
us to implement the phenomena discussed in Section~\ref{sec_results}, like the variation
of the TiO bands at short wavelengths as a function of C/O ratio and mass or the changes
caused by the CN features in warmer stars, in an approximated way. Needless to say that
the current interpolation routine is preliminary and may be modified significantly as we
expand the COMARCS grid.

The result of the interpolation are the bolometric corrections for any desired combination
of stellar parameters, which may be computed for a wide variety of photometric systems. This
allows us to convert the values tabulated in theoretical isochrones into magnitudes and
colours corresponding to the photospheres, if we study populations of stars. For the warmer
objects not covered by the COMARCS grid, we use other libraries, which are at the moment
either the one from \citet{2003IAUS..210P.A20C} as in \citet{2008A&A...482..883M} or the
PHOENIX BT-Settl data \citep{2012RSPTA.370.2765A,2013A&A...556A..15R} as in
\citet{2014MNRAS.444.2525C}. In the transition region, which occurs in the effective
temperature interval from 4000 to 5000~K, we make a linear interpolation between the
input sets. One can see in Fig.~\ref{ari_kuru} that in this range different hydrostatic
codes give very similar overall energy distributions. The plot shows low resolution spectra
based on COMARCS, MARCS \citep{2008A&A...486..951G} and Kurucz \citep{2003IAUS..210P.A20C}
models with 5000, 4500, 4000 and 3500~K having $\rm log(g~[cm/s^2]) = 2.0$ and the mass
as well as the abundances of the Sun. Considerable deviations, which are in general due
to the use of varying opacity data, appear almost only for the coolest atmospheres. Thus,
we expect in the region from 4000 to 5000~K a smooth transition between the photometric
results from different libraries.

When necessary, we consider the effect of the circumstellar dust shells on the derived
absolute magnitudes as described in \citet{2008A&A...482..883M}. This procedure is
currently being revised and improved (Nanni et al.\ in prep.). Theoretical isochrones
computed using transformations taken from the presented database are made available as
an additional option in the CMD web interface \url{http://stev.oapd.inaf.it/cmd}.

\section{Conclusions}
\label{sec_conclusions}

We have presented a grid of hydrostatic COMARCS atmospheres covering mainly stars with
effective temperatures up to 5000~K, which is intended to be used in combination with
stellar evolution calculations. The corresponding R~=~10000 opacity sampling and convolved
R~=~200 spectra as well as the bolometric corrections for a large number of photometric
systems can be downloaded from \url{http://starkey.astro.unipd.it/atm}. Since
the database is continuously growing, there will be frequent upgrades of the online
material in the future. For example, we recompute and extend the carbon star grid of
\citet{2009A&A...503..913A} considering the improvements described in Section~\ref{sec_opac}.
Also the amount of available models for S giants should be increased using s-process element
abundances predicted by the stellar evolution codes. We want to emphasize again that due to
their statistical nature the high resolution opacity sampling spectra must not be directly
compared to observations.

Using the data currently available in the COMARCS grid we have investigated the spectroscopic
and photometric properties of K and M giants. The consequences of adopting different masses,
C/O ratios and an increased nitrogen abundance are studied. In general, these quantities
cause much smaller changes than the main stellar parameters effective temperature, surface
gravity and overall metallicity. Nevertheless, in many cases they cannot be neglected depending
on the wavelength ranges of the filters and spectra and on the type of the star. A good example
is the phenomenon described in Section~\ref{sec_mass1} that the intensity of the TiO bands grows
with lower mass, while the molecular absorption in the mid and far infrared decreases. This
happens only in extended giants with $\rm log(g~[cm/s^2]) \le 0.0$ and changes the V, R and
I fluxes. In addition, the depth of the TiO features depends on the C/O ratio, which affects
also the lines of CO and water. However, in this case the behaviour is more complex, since
it varies with temperature. The trend appearing in the coolest models below about 3100~K that
the TiO bands get stronger with higher C/O values at least partly reverses in the warmer ones.
As one can see in Section~\ref{sec_phot} this is also reflected by the (V$-$K) and (V$-$I)
indices. The consequences of an enhancement of s-process elements often connected to an
increased C/O ratio are not discussed here, because we do not focus on S giants. A higher
nitrogen abundance, which has no effect below 3200~K, changes mainly the atomic and CN
features.

Models with variable masses, C/O ratios and nitrogen abundances are not available for all
combinations of the basic stellar parameters. In addition, we have no calculations covering a
simultaneous change of these quantities. Thus, their effect is included with correction
terms applied to the interpolation results, when the COMARCS grid is used to determine the
observable properties of stars. Those factors should be taken from COMARCS atmospheres having
parameters as close as possible, since they may vary a lot with temperature or surface
gravity. Therefore it is desirable to compute further sequences of models differing in
mass, C/O value and nitrogen abundance. Dense grids are at the moment only available for
a changed C/O ratio.

When we compare the photometric results obtained from the COMARCS grid to observed relations
and to data for the Galactic Bulge, we get in general a good agreement. Considerable deviations
appear only for the coolest stars with effective temperatures below about 3200~K at solar
metallicity and for very red colours of Mira variables. This is expected, since such giants
show strong pulsations and mass loss and may be obscured by circumstellar dust. Due to their
dynamical nature those phenomena cannot be covered by hydrostatic calculations. In that case
grids of models, which take the mentioned processes into account, are needed
\citep[e.g.][]{2014A&A...566A..95E,2015A&A...575A.105B}. For the combination with stellar
evolution results the influence of dust is often approximated with pure radiative transfer or
more simple stationary wind computations.

We have also investigated the effect of using different opacity data for water, which
create some deviations in the spectra of cool M giants. But the corresponding changes
of the photometric properties studied here are not large enough to select a preferred
line list, if one considers all other uncertainties appearing for such objects.

\section*{Acknowledgements}

This work was mainly supported by the ERC Consolidator Grant funding scheme
({\em project STARKEY}, G.A. n.~615604). PM acknowledges support from Progetto di Ateneo
2012, University of Padova, ID:~CPDA125588/12. Sincere thanks are given to M.A.T.~Groenewegen
and J.A.D.L.~Blommaert who provided mean OGLE-VI magnitudes for the Galactic Bulge Miras. This
research has made use of the NASA/IPAC Infrared Science Archive, which is operated by the Jet
Propulsion Laboratory, California Institute of Technology, under contract with the National
Aeronautics and Space Administration. This publication makes use of data products from the Two
Micron All Sky Survey, which is a joint project of the University of Massachusetts and the
Infrared Processing and Analysis Center/California Institute of Technology funded by the
National Aeronautics and Space Administration and the National Science Foundation. We thank
Marco Dussin for helping us with the electronic publication of the data.

\bibliographystyle{mnras}
\bibliography{msterne}

\appendix

\section{The Current COMARCS Grid}

\begin{table*}
\centering
\caption{The opacity sets available in the COMARCS grid. They are mainly characterized by the
abundances of the bulk of metals [Z/H] and of oxygen [O/H] relative to the solar values as well as
by the C/O ratio and microturbulent velocity ($\xi$). Additional parameters are listed in an
extra column. We have also included information about the lowest and highest effective temperature
($\rm T_{min,max}$) and $\rm log(g~[cm/s^2])$ covered by the models of a set. The status tells,
how many COMARCS atmospheres are available: grd = big grid, grh = small limited grid, seq = one or more
sequences, not = no models, only opacity tables.}
\label{ari_comarcs}
\begin{tabular}{rrrclccrrc}
\hline
[Z/H] & [O/H] & C/O & $\xi$~[km/s] & other parameters & $\rm T_{min}$~[K] & $\rm T_{max}$~[K] &
$\rm log(g)_{min}$ & $\rm log(g)_{max}$ & status\\
\hline
 0.00 &  0.00 &  0.550 & 2.5 &                                  & 2600 & 5940 & -1.00 & 5.32 & grd\\
 0.00 &  0.00 &  0.550 & 2.5 & BT2 water list                   &      &      &       &      & not\\
 0.00 &  0.00 &  0.550 & 2.5 & no CIA                           & 2600 & 5000 &  5.00 & 5.00 & seq\\
 0.00 &  0.00 &  0.550 & 1.5 &                                  & 2600 & 5800 &  0.00 & 5.00 & seq\\
 0.00 &  0.00 &  0.550 & 5.0 &                                  & 2600 & 5000 &  5.00 & 5.00 & seq\\
 0.00 &  0.00 &  0.550 & 2.5 & [N/Z]~=~+1.00                    & 2600 & 5000 &  0.00 & 2.00 & seq\\
 0.00 &  0.00 &  0.300 & 2.5 &                                  & 2600 & 4600 & -0.95 & 2.50 & grd\\
 0.00 &  0.00 &  0.700 & 2.5 &                                  & 2600 & 4000 & -0.95 & 2.00 & grd\\
 0.00 &  0.00 &  0.800 & 2.5 &                                  & 2600 & 3500 & -0.95 & 0.00 & grd\\
 0.00 &  0.00 &  0.900 & 2.5 &                                  & 2600 & 3500 & -1.00 & 0.00 & grd\\
 0.00 &  0.00 &  0.950 & 2.5 &                                  & 2600 & 3500 & -0.95 & 0.00 & grd\\
 0.00 &  0.00 &  0.970 & 2.5 &                                  & 2600 & 4000 &  0.00 & 0.00 & seq\\
 0.00 &  0.00 &  0.970 & 2.5 & [Y/Z] \& [Zr/Z]~=~+1.5           & 2500 & 4000 & -1.00 & 2.00 & grd\\
-0.41 & -0.41 &  0.550 & 2.5 &                                  &      &      &       &      & not\\
-0.41 & -0.64 &  0.374 & 2.5 & [O/Z]~=~$-0.23$, [N/Z]~=~+1.32   &      &      &       &      & not\\
-0.41 & -0.63 &  0.596 & 2.5 & [O/Z]~=~$-0.23$, [N/Z]~=~+1.32   &      &      &       &      & not\\
-0.41 & -0.63 &  0.841 & 2.5 & [O/Z]~=~$-0.23$, [N/Z]~=~+1.32   &      &      &       &      & not\\
-0.50 & -0.50 &  0.550 & 2.5 &                                  & 2600 & 5000 & -1.00 & 5.00 & grd\\
-0.50 &  0.00 &  0.275 & 2.5 & [O/Z]~=~+0.5, [alp/H]~=~Bulge    & 2600 & 5000 & -0.95 & 5.00 & grd\\
-0.70 & -0.70 &  0.550 & 2.5 &                                  & 2600 & 4500 &  0.00 & 2.50 & grd\\
-0.70 & -0.70 &  0.250 & 2.5 &                                  & 2600 & 4500 &  0.00 & 2.50 & grd\\
-1.00 & -1.00 &  0.550 & 2.5 &                                  & 2600 & 7171 & -1.00 & 5.00 & grd\\
-1.00 & -1.00 &  0.550 & 2.5 & no CIA                           & 2800 & 4000 &  5.00 & 5.00 & seq\\
-1.00 & -1.00 &  0.550 & 2.5 & [N/Z]~=~+1.00                    & 2600 & 7171 &  0.00 & 4.53 & seq\\
-1.49 & -1.49 &  0.550 & 2.5 &                                  & 2800 & 5000 &  4.50 & 5.00 & seq\\
-1.50 & -1.50 &  0.550 & 2.5 &                                  & 2600 & 5000 & -0.95 & 5.00 & grd\\
-2.00 & -2.00 &  0.550 & 2.5 &                                  & 2600 & 5000 & -1.00 & 5.00 & grd\\
 0.50 &  0.50 &  0.550 & 2.5 &                                  & 2600 & 5000 & -0.90 & 5.00 & grd\\
 1.00 &  1.00 &  0.550 & 2.5 &                                  & 2600 & 5000 & -0.65 & 5.00 & grd\\
 1.50 &  1.50 &  0.550 & 2.5 &                                  &      &      &       &      & not\\
 0.00 &  0.00 &  1.010 & 2.5 &                                  & 2600 & 4000 & -0.95 & 2.00 & grd\\
 0.00 &  0.00 &  1.010 & 2.5 & [Y/Z] \& [Zr/Z]~=~+1.0           & 2600 & 4000 &  0.00 & 0.00 & seq\\
 0.00 &  0.00 &  1.050 & 2.5 &                                  & 2500 & 4000 & -1.00 & 2.00 & grd\\
 0.00 &  0.00 &  1.100 & 2.5 &                                  & 2600 & 4500 & -0.95 & 2.00 & grd\\
 0.00 &  0.00 &  1.400 & 2.5 &                                  & 2600 & 4000 & -0.95 & 0.00 & seq\\
 0.00 &  0.00 &  2.000 & 2.5 &                                  & 2600 & 4200 &  0.00 & 0.00 & seq\\
 0.00 &  0.00 &  1.050 & 2.5 & [N/Z]~=~$-1.0$                   & 2600 & 4000 &  0.00 & 0.00 & seq\\
 0.00 &  0.00 &  1.050 & 2.5 & [N/Z]~=~$-5.0$                   & 2600 & 4000 &  0.00 & 0.00 & seq\\
 0.00 &  0.00 &  1.400 & 2.5 & [N/Z]~=~$-1.0$                   &      &      &       &      & not\\
 0.00 &  0.00 &  1.400 & 2.5 & [N/Z]~=~$-5.0$                   &      &      &       &      & not\\
-0.50 & -0.50 &  1.050 & 2.5 &                                  & 2600 & 4000 &  0.00 & 0.00 & seq\\
-0.50 & -0.50 &  1.100 & 2.5 &                                  & 2500 & 4000 &  0.00 & 0.00 & seq\\
-0.50 & -0.50 &  1.400 & 2.5 &                                  & 2500 & 4000 &  0.00 & 0.00 & seq\\
-0.50 & -0.50 &  2.000 & 2.5 &                                  & 2500 & 4000 &  0.00 & 0.00 & seq\\
-0.50 & -0.50 &  4.000 & 2.5 &                                  & 2500 & 4000 &  0.00 & 0.00 & seq\\
-0.50 & -0.50 &  8.000 & 2.5 &                                  & 2500 & 4000 &  0.00 & 0.00 & seq\\
-1.00 & -1.00 &  1.100 & 2.5 &                                  &      &      &       &      & not\\
-1.00 & -1.00 &  1.400 & 2.5 &                                  &      &      &       &      & not\\
-1.00 & -1.00 &  2.000 & 2.5 &                                  &      &      &       &      & not\\
-1.00 & -1.00 &  4.000 & 2.5 &                                  &      &      &       &      & not\\
-1.00 & -1.00 &  8.000 & 2.5 &                                  &      &      &       &      & not\\
-2.00 & -2.00 &  1.400 & 2.5 &                                  & 3600 & 4200 &  0.00 & 2.50 & grh\\
-2.00 & -2.00 &  2.000 & 2.5 &                                  & 3600 & 4200 &  0.00 & 2.50 & grh\\
-2.00 & -2.00 &  6.000 & 2.5 &                                  & 3600 & 4200 &  0.00 & 2.50 & grh\\
-2.00 & -2.00 & 10.000 & 2.5 &                                  & 3800 & 4200 &  0.00 & 2.50 & grh\\
-2.00 & -1.50 & 10.000 & 2.5 & [C/Z] \& [N/Z] \& [O/Z]~=~+0.5   & 3600 & 4200 &  0.50 & 2.50 & grh\\
\hline
\end{tabular}
\end{table*}

In Table~\ref{ari_comarcs} we list the various opacity sets that are currently available
within the COMARCS grid. Each of them is defined by a certain selection of elemental
abundances, microturbulent velocity and input data for the absorption and may include
model atmospheres with different effective temperature, surface gravity or mass. A description
of the corresponding calculations can be found in Section~\ref{sec_mods}. We want to
emphasize that the information in Table~\ref{ari_comarcs} represents only a snapshot
at the time when this paper was produced, since additional models will be computed in
the future. Thus, we do not give a complete list of available COMARCS atmospheres here.
The corresponding data may be obtained from
\url{http://starkey.astro.unipd.it/atm}.

The main parameters of the opacity sets are the overall metallicity [Z/H] as described in
Section~\ref{sec_par}, the abundance of oxygen [O/H], the C/O ratio and the microturbulent
velocity. We have also varied the amount of some key elements like N or Zr. [alp/H]~=~Bulge
in Table~\ref{ari_comarcs} stands for the $\alpha$ abundances from
\citet{2010A&A...512A..41B,2011A&A...533A.134B}. In addition, we have computed dwarf
models without collision induced absorption. The quantities [Z/H], [O/H] and C/O are
given in the listings and files of the database by the number density of iron, oxygen
and carbon relative to hydrogen $\rm log(\varepsilon_{Fe,O,C}/\varepsilon_H) + 12$.

\section{The Excess of Free Oxygen}

\begin{figure}
\includegraphics[width=7.0cm,clip,angle=270]{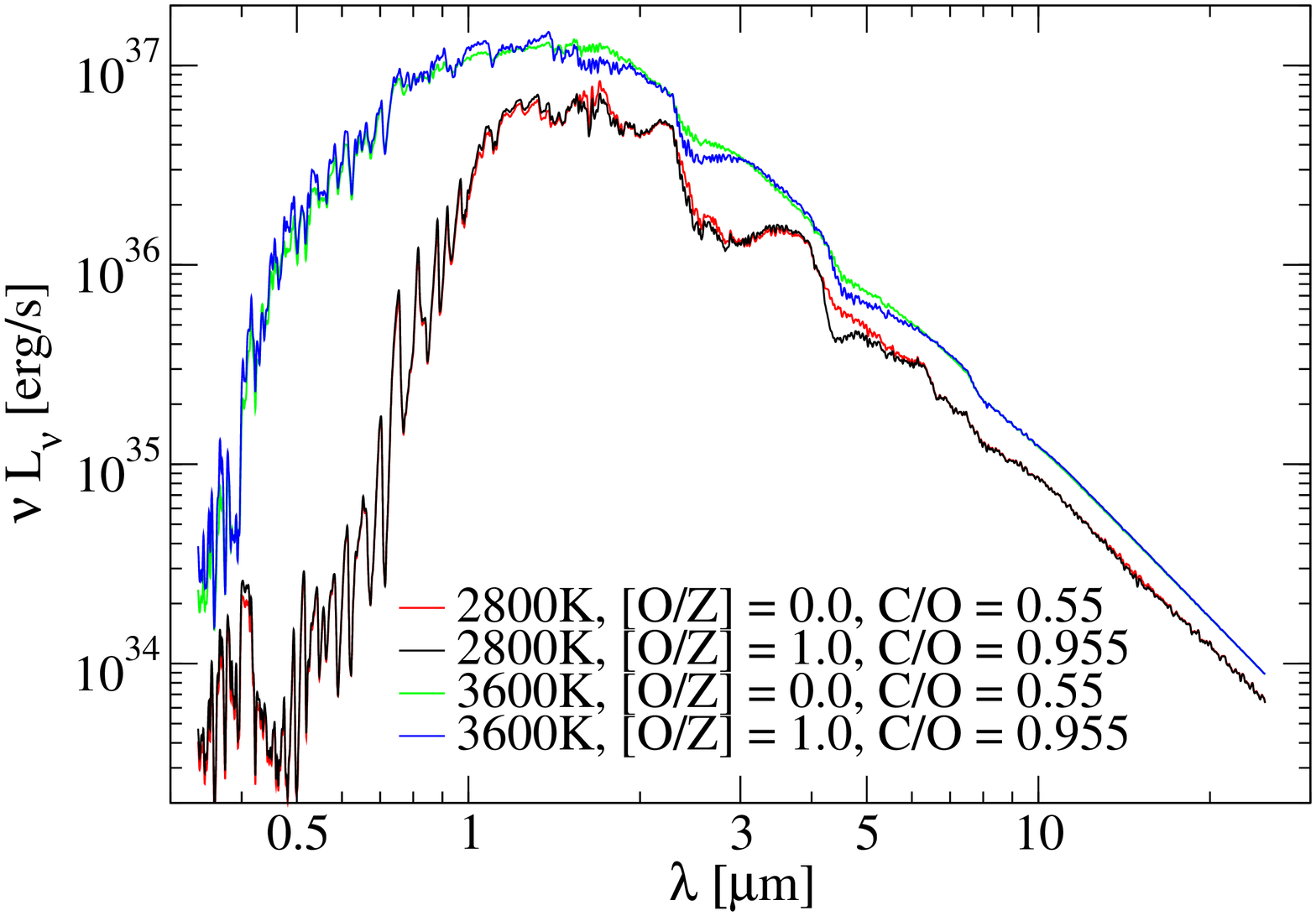}
\caption{Spectra based on COMARCS models with $\rm T_{eff} = 2800~K$ or
3600~K, $\rm log(g~[cm/s^2]) = 0.0$, one solar mass and solar metallicity. The effect
of changing [O/Z] from 0 to +1 at a constant (O$-$C) of 8.455 is shown.}
\label{ari_cminuso}
\end{figure}

Since in the cooler regions of an M star atmosphere most of the carbon atoms form CO
molecules, only the remaining amount of oxygen can be used to create other species like
TiO, VO, H$_2$O or SiO, which represent important absorbers. Thus, the excess (O$-$C)
given by $\rm log((\varepsilon_O - \varepsilon_C)/\varepsilon_H) + 12$ may dominate the
overall opacity, if the abundances of C and O are changed. In order to investigate the
effect of this quantity we have calculated COMARCS models with [O/Z] increased to +1
keeping the (O$-$C) value of the Sun constant at 8.455. For the adopted solar metallicity
that results in a C/O ratio of 0.955 instead of 0.55 at [O/Z]~=~0. A comparison of spectra
for giants with 2800 and 3600~K having normal and enhanced oxygen abundances is shown in
Fig.~\ref{ari_cminuso}.

For the cooler effective temperature the spectra with [O/Z]~=~0 and +1 are very similar,
which is also true for the atmospheric structures. Differences appear mainly in the regions
of the CO bands with the largest changes for the ground state, because the amount of this
species increases at higher oxygen abundances. As expected, the absorption produced by the
other molecules shows almost no variation, if (O$-$C) remains constant. For the hotter
models the situation is more complex. In addition to the CO bands, which become again deeper
at higher [O/Z], several spectral features and the atmospheric structures change moderately.
Even if the abundances of important metals like Ti or V are kept constant, (O$-$C) is in
contrast to the cooler giants not the only crucial parameter. This may be explained by the
fact that in warmer stars the relative contribution of CO to the total opacity is larger and
a significant fraction of the carbon atoms can form CN or CH.

\bsp
\label{lastpage}
\end{document}